\documentclass[12pt]{article}

\usepackage{amsmath,amssymb,graphicx,longtable}
\usepackage{hyperref}

\usepackage{multicol,color}
\definecolor{darkred}{rgb}{0.5,0,0.5}
\definecolor{darkgreen}{rgb}{0,.5,0}
\hypersetup{pdfborder={0 0 0},colorlinks=true,urlcolor=darkred,citecolor=blue,linkcolor=darkgreen}

\makeatletter

\@addtoreset{equation}{section}
\makeatother
\newcommand{\be}{\begin{eqnarray}}
\newcommand{\ee}{\end{eqnarray}}
\newcommand{\nn}{\nonumber}

\newcommand{\mc}[1]{{\mathcal{#1}}}
\newcommand{\mf}[1]{{\mathfrak{#1}}}

\newcommand{\cL}[2]{\stackrel{(#1)}{\mathfrak{L}}{}^{\!\! #2}}
\newcommand{\J}[2]{\stackrel{(#1)}{J}{}^{\!\! #2}}
\newcommand{\cLtxt}[2]{\mathfrak{L}^{(#1)\, #2}}

\newcommand{\Lb}[2]{\stackrel{(#1)}{\mathfrak{C}}{}^{\!\! #2}}
\newcommand{\Jb}[2]{\stackrel{(#1)}{I}{}^{\!\! #2}}

\newcommand{\lb}{\left[}
\newcommand{\rb}{\right]}
\newcommand{\ints}{\mathbb{Z}}
\newcommand{\reals}{\mathbb{R}}

\begin{document}
{\flushright
{IHES/P/09/47}\\
{ULB-TH/09-39}\\
{AEI-2009-112}\\}

\vskip .7 cm
\hfill
\vspace{10pt}
\begin{center}
{\large {\bf Sugawara-type constraints in hyperbolic coset models}}

\vspace{1.1cm}

{\centering \rule[0.1in]{13cm}{0.5mm} }

\vspace{.5cm}

Thibault Damour\footnotemark[1], Axel Kleinschmidt\footnotemark[2] and Hermann Nicolai\footnotemark[3]

\vspace{10pt}
\footnotemark[1]{\em Institut des Hautes Etudes Scientifiques\\
35, Route de Chartres, FR-91440 Bures-sur-Yvette, France}\\[5mm]

\footnotemark[2]{\em Physique Th\'eorique et Math\'ematique\\
Universit\'e Libre
de Bruxelles \& International Solvay Institutes\\ 
ULB-Campus Plaine C.P. 231, BE-1050 Bruxelles, Belgium}\\[5mm]

\footnotemark[3]{\em Max-Planck-Insitut f\"ur Gravitationsphysik, Albert-Einstein-Institut\\
Am M\"uhlenberg 1, DE-14476 Potsdam, Germany}\\[10mm]

\end{center}

\vspace{3pt}
\begin{center}
\textbf{Abstract}\\[5mm]
\parbox{13cm}{\footnotesize
In the conjectured correspondence between supergravity and geodesic models on 
infinite-dimensional hyperbolic coset spaces, and $E_{10}/K(E_{10})$ in 
particular, the constraints play a central role. We present a Sugawara-type 
construction in terms of the $E_{10}$ Noether charges that extends these 
constraints infinitely into the hyperbolic algebra, in contrast 
to the truncated expressions obtained in 
\href{http://arxiv.org/abs/0709.2691}{arXiv:0709.2691} that involved only 
finitely many generators. Our extended constraints are associated to an infinite
set of roots which are all imaginary, and in fact fill the closed past
light-cone of the Lorentzian root lattice.
The construction makes crucial use of the $E_{10}$ Weyl 
group and of the fact that the $E_{10}$ model contains  both $D=11$ 
supergravity and $D=10$ IIB supergravity. Our extended constraints appear to unite
in a remarkable manner the different canonical 
constraints of these two theories. This construction may also shed new light
on the issue of `open constraint algebras' in traditional canonical 
approaches to  gravity.}

\end{center}

\thispagestyle{empty}
\newpage
\setcounter{page}{1}

\begin{section}{Introduction}

In canonical formulations of gravity, the constraints are the essential 
ingredients for, and main obstacles to, carrying out a canonical 
quantization of gravity~\cite{DeWitt:1967yk} (for an overview and 
bibliography see~\cite{Kiefer:2004gr}). This applies in particular
to the Hamiltonian (scalar) constraint determining evolution in `time', and 
therefore the dynamics. The problem of properly setting up and defining 
the quantum constraints has been tackled in a variety of approaches but, 
arguably, the problem remains as open as in Bryce DeWitt's seminal 1967 
paper~\cite{DeWitt:1967yk}. A further cause of difficulties, shared by all
approaches so far, can be traced to the fact that the constraints form 
an {\em open algebra}, that is, the structure `constants' are not 
constants, but field dependent. 

At the level of classical maximal supergravity, progress has been made in 
the last years towards establishing a correspondence between the equations 
of $D=11$ supergravity on the one hand and a geodesic coset model based on 
the hyperbolic Kac--Moody structure $E_{10}$~\cite{Damour:2002cu} 
on the other (similar correspondences exist for other supergravity models).
The supergravity equations are treated canonically and therefore comprise 
{\em dynamical} (evolution) equations and {\em constraint} equations.  There is a 
precise correspondence between a truncation of the {\em dynamical equations} 
and a truncation of the {\em geodesic equation} on the coset $E_{10}/K(E_{10})$~\cite{Damour:2002cu}. 
The $D=11$ supergravity {\em constraint} equations can similarly 
be mapped to {\em constraints} that can be imposed consistently on the 
geodesic motion~\cite{Damour:2007dt}. For instance, imposition of the 
Hamiltonian constraint implies that the geodesic is null. According
to~\cite{Damour:2007dt} the weakly conserved constraints of $D=11$ 
supergravity can be translated into weakly conserved coset model 
constraints, which in turn allow for a reformulation as bilinear expressions 
in terms of conserved charges, that is, as {\em strongly} conserved
constraints.\footnote{As usual, the term `weakly conserved constraints' 
here refers to a set of constraints $\mc{C}$ satisfying (modulo the coset 
equations of motion) $d\mc{C}/dt = f(\mc{C})\approx 0$,
where $f(\mc{C})$ is a function vanishing on the {\em constraint surface} defined by $\mc{C}=0$, while
`strongly conserved' constraints satisfy $d\mc{C}/dt=0$ 
(upon use of the equations of motion).}
As noted there, this construction is very reminiscent of the well-known 
Sugawara construction~\cite{Sugawara:1967rw} for affine Lie 
algebras~\cite{Bardakci:1970nb,Goddard:1986bp}. It is the purpose of 
the present paper to follow up on this observation, making it more 
precise and giving the beginning of a generalized Sugawara construction 
for hyperbolic Kac--Moody algebras which makes the analogy
with the affine construction much more compelling.

Understanding and reformulating supergravity in these algebraic terms could 
prove very useful for the transition to the quantum theory (see \cite{KKN}
for first steps towards the quantization of the $E_{10}/K(E_{10})$ model and \cite{Forte:2008jr} for pure gravity).
An analogy to be kept in mind in this discussion is that of (bosonic) string 
theory. There, the dynamical equation for the embedding (target space) 
coordinates can be written as a free wave equation if one adopts a conformal 
gauge. This free wave equation admits an infinite set of conserved charges
$ \alpha_n^{\mu}$. The price 
to pay for the simple dynamical equation is that one has to impose the 
(Fubini-Veneziano-)Virasoro constraints, $ L \sim \alpha \, \alpha$,
on the solutions. In the quantum version, the Virasoro constraints and the existence of a proper Hilbert space imply the critical dimension~\cite{Goddard:1972iy}. Assuming the validity of the Kac--Moody/supergravity correspondence, the dynamical equations of supergravity also become simple, 
yielding geodesics on a symmetric space as their solutions. This system is fully integrable.
It admits an infinite set of conserved charges,  $J$, that do not (Poisson) commute 
among themselves, and one can formally write down the general solution in terms of
$J$ and some initial data. The complications and interesting 
structures are then again to be found in the constraints and their algebra. The fact that 
all constraints found so far admit a Sugawara-like structure,  {\it i.e.}, $ \mf{L} \sim J \, J$,
is tantalizing in this analogy, and may turn out to 
be crucial for the quantisation of the theory. The gauge symmetries encoded in the coset 
constraints are directly linked to the space-time and gauge symmetries that are known 
from the geometrical formulation of supergravity. 

The replacement of the supergravity constraints by coset model constraints 
with an underlying algebraic structure may also shed new light on the old 
problem of open constraint algebras alluded to above, circumventing some of 
the seemingly insurmountable difficulties of the usual canonical formulation. 
The main new feature here is that the `structure constants', while still 
dependent on the dynamical degrees of freedom (fields), become 
{\em constants of motion} in the present formulation. More explicitly, 
suppose the classical constraints $\mc{C}^A(\phi)$ satisfy the first-class 
canonical (Poisson) algebra
\be
\left\{ \mc{C}^A(\phi), \mc{C}^B(\phi)\right\} = 
f^{AB}{}_C(\phi) \mc{C}^C(\phi)\,,
\ee
where $\phi$ denotes the canonical variables. In the standard formulation
of canonical gravity and supergravity, the $\phi$-dependent structure 
`constants' $f^{AB}{}_C(\phi)$ do not (Poisson) commute with the 
Hamiltonian and thus vary in time. By contrast, the structure constants 
obtained with the Sugawara-like form of the constraints do commute 
with the Hamiltonian constraint, and are thus preserved 
in time, even though they still depend on the canonical variables $\phi$.
Because the correspondence between the space-time based field theory and 
the one-dimensional $E_{10}/K(E_{10})$ model is only very incompletely 
understood, it is, however,  not clear how to translate the coset model 
constraints back into more conventional field theory language. At the 
very least, one can say that the relation  between the field variables 
of the geometric theory and the $E_{10}$ variables must be extremely non-local.

Obtaining a universal algebraic description of the constraints and their 
algebra is also desirable from an M-theory point of view. In the same way 
that the unique dynamical geodesic equation on $E_{10}/K(E_{10})$ allows for maps 
to {\em different} maximal supergravity theories, depending on the level 
decomposition chosen to describe the infinite-dimensional Lie 
algebra~\cite{Kleinschmidt:2004dy,Damour:2004zy,Kleinschmidt:2004rg,Henneaux:2008nr}, the constraints should also exhibit this `versatility'. Our 
construction below has this property, albeit in a novel way. More precisely, 
we will define a `universal scaffold' of hyperbolic Sugawara 
constraints by using null root vectors $\alpha$ of the hyperbolic algebra, 
decomposed into a sum of two real roots 
$\beta_1+\beta_2=\alpha$,  and the hyperbolic Weyl group. This will define 
an infinite number of constraints $\mf{L}_\alpha$ associated with a
`skeleton' of roots $\alpha$
{\em on} the light-cone  in terms of current 
bilinears. (The notions of skeleton and scaffold are depicted in figures~\ref{fig:skeleton} and~\ref{fig:scaffold} below.) Extending (away from the real $\beta_i$ case) the set of current-bilinear contributions
$\mf{L}_\alpha\sim J_{\beta_1} \,J_{\beta_2}$ to a given null-root constraint  ($\alpha^2=0$),
or extending the skeleton of supporting roots $\alpha$
constraints {\em into} the light-cone ($\alpha^2 <0$), however, seems to
require the choice of a subalgebra of the hyperbolic algebra that is kept 
manifest. In analogy with affine algebras, this procedure is very suggestive 
of a choice of `spectral parameters' for the hyperbolic algebra, even 
though we do not know whether such a realization of the hyperbolic
algebra exists. However, the picture that emerges from the present
work is that if such realizations exist, they do so only in combination
with suitable constraints. Furthermore, such realizations cannot 
be unique, giving the algebra a `chameleon-like' aspect. This feature 
would be in line with the conjectured emergence of a space-time structure 
from the Lie algebra, where the dimension of the emergent space would 
depend on the decomposition and the chosen form of the constraints,
such that the `spectral parameters' would become associated to spatial
coordinates\footnote{However, this association is likely to be more subtle 
  than just a simple equality, as can already be seen for the affine 
  spectral parameter in $D=2$ supergravities, cf. Eqn.~(2.1) of \cite{NS}
  with $\rho=t$ (time) and $\tilde\rho = x^1$ (space).}. These points 
will be elaborated on and explained below by means of the constraints 
of $D=11$ supergravity and of type IIB supergravity, respectively, but
similar results are expected to hold for other decompositions, such as
massive IIA theory, as well as for maximal supergravities in lower 
dimensions. Importantly, though the set of roots `supporting' the constraints
is clearly related to the weight diagram of particular highest-weight
representations of $E_{10}$, the constraints themselves 
do {\em not} form (under Poisson commutation) a highest or lowest weight representation of the 
hyperbolic $E_{10}$, as already observed in~\cite{Damour:2007dt}, and 
explained in  much more detail here. Rather, they indicate the existence 
of new unexplored algebraic structures inside the hyperbolic algebra 
and its envelopping algebra.  

We emphasize that our approach is canonical and crucially relies on a 
split of space and time, as well as certain gauge choices required for 
matching the supergravity and coset model degrees of freedom. An 
earlier and conceptually different M-theory proposal based on the indefinite, 
but non-hyperbolic, `very extended' Kac--Moody algebra $E_{11}$ has been 
developed by Peter West and collaborators~\cite{West:2001as,West2003}. 
In contradistinction to the present work, their approach is `covariant' in 
the sense that neither a split of space-time nor gauge choices for the 
supergravity fields are required, and the issue of writing down canonical 
constraints thus does not arise in the same way. Instead, one needs 
to introduce extra gauge invariances encompassing the gauge transformations 
of supergravity, and the problem becomes one of `fitting' such gauge 
symmetries into the $E_{11}$ framework~\cite{Riccioni:2009hi}.
However, despite many similarities at the kinematical level, especially 
with regard to embedding the bosonic sectors of maximal supergravities~\cite{Schnakenburg:2001ya,Schnakenburg:2002xx,Kleinschmidt:2003mf,West2004}, 
it appears doubtful whether a gauge-fixed version of that approach 
matches with the structures presented here.

{}From the mathematical point of view, it would also be desirable to associate 
a Sugawara-type construction to a hyperbolic algebra. In the affine case, 
the existence of this construction is directly linked to the realization 
of affine algebras as loop algebras via the so-called spectral parameter.
A similar description and understanding is lacking for hyperbolic algebras, 
the only known description is in terms of generators and relations in the 
Chevalley--Serre basis. Any construction hinting at an alternative description
could shed light on the deeper and to date elusive structure of hyperbolic 
Kac--Moody algebras. After all, even not knowing about the current algebra 
realization of affine algebras, the existence of a preferred set of 
bilinear Virasoro operators in the envelopping algebra would almost
inevitably lead to this realization. Here, we are searching for a similarly 
distinguished structure in the envelopping algebra of the hyperbolic algebra.

The remainder of the paper is structured as follows. In section~\ref{sugcon} 
we first review the affine Sugawara construction and rephrase it in a 
slightly unconventional form. We use this form to propose a (partly 
schematic) trial expression for Sugawara generators for hyperbolic algebras. 
In section~\ref{Univsec} we then explore this trial expression in more 
detail in the case of $E_{10}$ and show that our trial expression does not only serve to reproduce 
the $D=11$ constraints but also those of type IIB supergravity. This also 
allows for a more precise definition of the Sugawara constraints and an 
exploration of their structure in terms of a skeleton of constraints 
associated with null roots and terms induced by covariantization. In 
appendices, we collect some known results on level decomposition in order 
to render the presentation self-contained, as well as some more detailed computations.

\end{section}

\begin{section}{Sugawara construction}
\label{sugcon}

Before proceeding to the discussion of the hyperbolic Sugawara construction 
we first review briefly the definition of Sugawara operators for affine Lie 
algebras, see~\cite{Goddard:1986bp} (as well as \cite{Sugawara:1967rw,Bardakci:1970nb} for earlier 
work).

\begin{subsection}{Affine Sugawara construction}

 A non-twisted affine Lie algebra can be defined for any 
finite-dimensional Lie algebra. Let the finite-dimensional Lie algebra 
$\mf{g}$ be simple and generated by $T^A$ ($A=1,\ldots,\dim\mf{g}$) with 
commutation relations $\lb T^A, T^B\rb = f^{AB}{}_C T^C$ and non-degenerate 
invariant form $\langle T^A|T^B\rangle =\kappa^{AB}$. Then the corresponding 
affine Lie algebra $\hat{\mf{g}}$ has generators $T_m^A$ (for $m\in\ints$), $c$
and $d$ with non-trivial commutation relations
\be\label{affcom}
\lb T^A_m, T^B_n \rb = f^{AB}{}_C T^{C}_{m+n} + 
\kappa^{AB} m \delta_{m,-n} c\,,\quad 
\lb d,T^A_m\rb = -m T^A_m\,.
\ee
The generator $c$ commutes with all Lie algebra generators and is called 
the {\em central element},\footnote{The central element of the affine Lie algebra,
here denoted $c$, is often denoted $K$; it should not be confused with the central
element of the Virasoro algebra associated to the affine algebra.} 
while the generator $d$ is called the 
{\em derivation}.\footnote{This terminology follows from the presentation 
 of affine algebras as loop algebras where $d$ is the derivative with 
 respect to the spectral parameter~\cite{Goddard:1986bp}.} 

In any irreducible highest weight representation, the central element $c$
acts as a scalar; its eigenvalue $k$ on that representation is called the 
{\em level} of the representation. For such a level $k$ representation, the 
Sugawara generators are defined (within the enveloping algebra of the $T^A_m$'s) by~\cite{Goddard:1986bp} (for $n\in\ints$) 
\be\label{sugaff}
L_n = \frac{1}{2(k+h^\vee)} \sum_{m\in\ints} : T^A_{n-m} T^B_m: \kappa_{AB}\,,
\ee
where the colons denote normal ordering as appropriate for the highest 
weight representation and $\kappa_{AB}$ is the inverse of $\kappa^{AB}$; $h^\vee$ is the dual Coxeter number defined by $f^{AC}{}_D f^{BD}{}_{C}=2h^\vee \kappa^{AB}$. We note that there are two separate contributions to the normalization of the Sugawara generators (\ref{sugaff}): The first one is $k$, related to the central extension, the second one $h^\vee$ comes from normal ordering. Both  contributions are {\em quantum effects}. Below, we will treat these two contributions differently. In the hyperbolic extension, the central generator  ceases to be central and is on par with all the other Lie algebra generators. Normal ordering, on the other hand, will be mostly ignored, as  our discussion deals with the classical constraints only.
Normal ordering ensures that the generators $L_m$ are well defined on any element 
of the representation. The operators (\ref{sugaff}) obey a Virasoro algebra 
\be\label{vircentral}
\lb L_m, L_n \rb = (m-n)L_{m+n} +\frac{k\,\dim\mf{g}}{12(k+h^\vee)}
m (m^2-1) \delta_{m,-n}\,
\ee
Their commutators with the affine generators are 
\be\label{LmTn}
\lb L_m, T^A_n \rb = -n T^A_{m+n}\,.
\ee

Here, we would like to take a more formal point of view and rewrite 
(\ref{sugaff}) as a quadratic expression in the generators without 
resorting to an integrable representation. The reason is that the 
normalization in (\ref{sugaff}) involves the {\em inverse} of the (shifted) 
eigenvalue of the central generator $c$. However, in the full hyperbolic 
algebra the element $c$ is no longer central (in fact, the hyperbolic
algebra does not possess any central elements), and a direct generalization
of (\ref{sugaff}) would thus necessarily involve the inverse of an 
operator, which furthermore is no longer singled out in the full
algebra. For this reason, we formally multiply (\ref{sugaff}) by the 
central element and drop the normalization constant. We also 
recall that affine Lie algebras have two different kinds of roots: {\em real} 
roots and {\em null} roots. In particular, there is a primitive null 
root $\delta$ which can be used to describe all roots of the affine 
algebra via an affine  ladder diagram: Let $\Delta^{\text{fin}}\equiv 
\Delta(\mf{g})$ be the set of roots of the finite-dimensional algebra 
$\mf{g}$ (where we include $\alpha=0$ for simplicity), then the root 
system of the affine extension $\hat{\mf{g}}$ is 
\be\label{deltaaff}
\Delta^{\text{aff}} \equiv \Delta(\hat{\mf{g}}) = 
\left\{\alpha+n\delta\,:\, \alpha \in \Delta^{\text{fin}} \text{ and } n\in\ints\right\}\,,
\ee
that is, there are $\ints$ copies of the finite root system. The roots 
$n\delta$ are null roots and the associated root space 
$\hat{\mf{g}}_{n\delta}$ has dimension given by the rank: 
$\text{mult}(n\delta) = \dim \hat{\mf{g}}_{n\delta} = \text{rank}(\mf{g})$ 
for $n\neq 0$. For $n=0$ the dimension is equal to that of the Cartan 
subalgebra and takes the value $\text{rank}(\mf{g})+2$ (the two extra
elements are $c$ and $d$). All other roots are real and the 
corresponding root spaces are one-dimensional. 

Using the structure of the affine root system we can rewrite the commutation relations (\ref{affcom}) as
\be
\lb T_{\alpha_1}, T_{\alpha_2} \rb = f_{\alpha_1\,\alpha_2}{}^{\alpha_1+\alpha_2} T_{\alpha_1+\alpha_2} + \kappa_{\alpha_1,\alpha_2} c\,,
\ee
where we have suppressed the multiplicity index for null roots. The values of $f_{\alpha_1\alpha_2}{}^{\alpha_1+\alpha_2}$ and $\kappa_{\alpha_1,\alpha_2}$ can be obtained by comparison with (\ref{affcom}). We furthermore define quadratic generators in the enveloping algebra $U(\hat{\mf{g}})$ by
\be\label{sugaff2}
L_{n\delta} := \sum_{\beta\in\Delta^{\text{aff}}} 
T_{n\delta-\beta} T_{\beta}\,,
\ee
where $T_\beta$ is a canonically normalized element in the root space $\hat{\mf{g}}_\beta$. If 
the $\beta$ root space is degenerate, we choose an orthonormal basis and contract with the canonically conjugate basis. Then 
the definition (\ref{sugaff2}) is unambiguous except when the root spaces 
of $n\delta-\beta$ and $\beta$ have different dimensions. This happens 
{\em only} when one of $n\delta-\beta$ or $\beta$ is equal to zero, {\it i.e.}, when 
one of the generators belongs to the Cartan subalgebra. In that case the 
generators are to be contracted according to the definition (\ref{sugaff}), 
{\it i.e.}, we omit any terms involving a contraction with $c$ or $d$, but contract 
only with elements of the Cartan subalgebra of the horizontal $\mf{g}$. Except 
for this point and the lack of normal ordering, the expression (\ref{sugaff2})
is a reformulation of (\ref{sugaff}). Note that although we could have 
defined quadratic generators of the form (\ref{sugaff2}) for any point on 
the root lattice, we do this only for {\em null roots}. To get a Virasoro 
algebra it is furthermore essential that the space of null roots has an 
additive structure since all null roots lie on a $\ints$-graded line.

The affine Weyl group is the semi-direct  product of the finite Weyl group 
with a translation group \cite{Kac}. After the standard embedding of the affine algebra into a hyperbolic algebra of over-extended type~\cite{Damour:2002fz}, the affine Weyl group can also be described as the subgroup of the hyperbolic Weyl group stabilizing an affine null root \cite{Gebert:1996vv};
the so-called affine translations are then realized as Lorentz boosts 
along this null direction.\footnote{We also note that the Weyl orbit of the `cusp' $\delta$ 
is dense on the boundary of the hyperbolic space obtained by projecting 
the interior of the forward lightcone onto the unit 
hyperboloid. Equivalently, the rays through all the 
 hyperbolic null roots cover the boundary of the lightcone densely.}
 Since null roots $n\delta$ are stabilized by the affine Weyl group $\mc{W}^{\text{aff}}$, 
 the l.h.s. of the definition  (\ref{sugaff2}) is invariant under the action of the 
 Weyl group. One can check that the r.h.s. is also invariant.

Besides the convention for null root spaces, the definition (\ref{sugaff2}) 
differs from the standard one (\ref{sugaff}) by its lack of normal ordering. 
However, as is well known, this affects only the generator $L_0$ for affine 
algebras. In addition, normal ordering is only required for the quantum 
theory, whereas we are here mainly concerned with the structure of the 
{\em classical} constraints. In the classical theory, one associates to 
each symmetry generator $T_{\alpha}$ a corresponding conserved charge,
say $J_{\alpha}$. Accordingly, we will below consider expressions such as 
(\ref{sugaff2}) (with the replacement $T_{\alpha} \to J_{\alpha}$) as functions on 
phase space and leave open the quantum definition of the constraints. 
We also remark that the generator $L_0$ as defined in (\ref{sugaff}) differs 
from the Hamiltonian (quadratic Casimir) by a term proportional to $c d$. 
Omission of this term is admissible in the affine case, but not in the
hyperbolic algebra. [In other words, our hyperbolic-algebra generalization of
(\ref{sugaff}) will contain terms of the type $c d$, which do not enter the
affine version of (\ref{sugaff}).] Correlatively, while the affine Hamiltonian is bounded below, {\it i.e.},
$L_0\geq 0$, the full Hamiltonian is not because the Cartan-Killing 
metric on the Cartan subalgebra is indefinite for hyperbolic algebras 
(with $\langle c | c \rangle = \langle d | d\rangle =0$ and 
$\langle c|d\rangle =1$).

We now proceed to compute the algebra of the constraints as defined by (\ref{sugaff2}). In the course of the following computations we manipulate infinite sums formally, well aware that they are not well-defined and normally would require a normal ordered evaluation on a representation space. With this in mind one computes in the universal enveloping algebra
\be
\lb L_{m\delta}, T_{\alpha} \rb 
&=& -2 \kappa_{\alpha,-\alpha}  c\,T_{m\delta+\alpha}\,.
\ee
which is he same as  (\ref{LmTn}), but now expressed in terms of affine roots.
The important point we wish to emphasize here is that the r.h.s is {\em bilinear} in affine generators since we multiplied the Sugawara generators by the central element. Continuing now to the commutator of two Sugawara generators (\ref{sugaff2}) leads to
\be\label{affvir}
\lb L_{m\delta}, L_{n\delta} \rb
&=& 2(m-n)\, c \, L_{(m+n)\delta}\,,
\ee
so that in this formulation the algebra closes with a pre-factor ($=c$) that 
is itself an algebra generator. Due to the lack of normal ordering one does 
not obtain the central term as in (\ref{vircentral}). Neither is the shift by 
the dual Coxeter number visible in this formal computation in the enveloping 
algebra.

\end{subsection}

\begin{subsection}{Hyperbolic Sugawara construction}

The expression (\ref{sugaff2}) can be formally generalized to hyperbolic 
Lie algebras of the over-extended type~\cite{Damour:2002fz}.\footnote{By 
  `over-extension' we mean the canonical extension via the non-twisted affine extension, 
  whereby two nodes are added to the Dynkin diagram; adding a third node 
  would yield `very-extended' algebras \cite{Gaberdiel:2002db}.}
In the hyperbolic case the root system $\Delta^{\text{hyp}}$ is much more 
complicated than (\ref{deltaaff}): Besides the real and null roots there 
are now time-like (purely imaginary) roots $\alpha$ with $\alpha^2<0$. The 
multiplicities of these roots grows exponentially and no closed formula for 
their multiplicities is known although these can be computed algorithmically, for example
via the Peterson recursion formula. For each $\alpha$ root space
$\mf{g_{\alpha}} \subset \mf{g}$ , we choose 
a basis
\be
T_\alpha^{(s)} \quad\text{ for $s=1,\ldots,\text{mult}(\alpha)$}.
\ee
which is `null orthonormal' (when using the standard bilinear form) 
with respect to the corresponding dual basis in the $\mf{g_{-\alpha}}$ root space:
\be\label{hypcr}
\langle T_\alpha^{(s)} | T_{\beta}^{(s')} \rangle = \delta_{s,s'}
\delta_{\alpha+\beta,0} \,.
\ee
The commutation relations are then
\be
\lb T_{\alpha_1}^{(s_1)}, T_{\alpha_2}^{(s_2)} \rb = f_{\,\,\,\alpha_1\,\,\,\alpha_2\,\,\, \,(s_{12})}^{(s_1)(s_2)\,\alpha_1+\alpha_2} T_{\alpha_1+\alpha_2}^{(s_{12})}\,.
\ee

Our hyperbolic generalization of the affine Sugawara construction (\ref{sugaff2})
then consists of two elements:
\begin{itemize}
 \item[(i)] the choice of a special set of `constraint' generators,
labelled by a subset, say 
$ \mc{C}$, of the set of pairs $(\alpha,\bar{s})$ labelling the roots (including their
degeneracy); and 

\item[(ii)] a general
 expression for the hyperbolic Sugawara generator $\mf{L}_{\alpha, \bar{s}}$
 (or `generalized Virasoro constraint')
 associated to a particular pair\footnote{Note that while $\alpha$ runs over a subset of 
 $\Delta$, $\bar{s}$ correspondingly runs over a subset of the full degeneracy of
 the root $\alpha \in \Delta$.} $(\alpha,\bar{s}) \in \mc{C}$  of the form
\be\label{sughyp}
\mf{L}_{\alpha, \bar{s}} = \sum _{\beta_1,\beta_2\in\Delta^{\text{hyp}}\atop \beta_1+\beta_2=\alpha} \sum_{s_1,s_2}
M_{s_1,s_2}(\beta_1,\beta_2)
T^{(s_1)}_{\beta_1} T^{(s_2)}_{\beta_2} \,.
\ee
\end{itemize}
Here $M_{s_1,s_2}(\beta_1,\beta_2)$ denote some numerical coefficients
that we expect to be simply $\pm 1$ or $0$ (or possibly other rational numbers) 
for an appropriate choice of the dual bases $T_{\pm \alpha}^{(s)}$
in the $\pm \alpha$ root spaces. 

We do not have yet a full understanding of the precise set $\mc{C}$ of `constraint'
generators\footnote{The letter $\mc{C}$ is used here to evoke both the word `constraint',
and the fact that the set $\mc{C}$ appears to have the structure of a convex cone.}, nor 
of the numerical coefficients  $M_{s_1,s_2}(\beta_1,\beta_2)$ entering
the definition of our generalized Virasoro constraints $\mf{L}_{\alpha, \bar{s}}$.
We will argue that 
a distinguished role is played by the `null subset' of  $\mc{C}$, {\it i.e.}, by the
case where  $\alpha$ is a null root. In that case, the corresponding
constraint degeneracy index takes only one value (while the degeneracy of a null root
within the hyperbolic algebra is equal to the rank).
Moreover, still in the case where $\alpha$ is a null root, we will be able to verify
that the coefficients $M_{s_1,s_2}(\beta_1,\beta_2)$ in (\ref{sughyp}) are indeed simply equal
to $\pm 1$ when both of $\beta_1$ and $\beta_2$ (such that $\alpha=\beta_1+\beta_2$)
are {\it real} roots. In the following, we shall refer to the better understood
`null' subset of $\mc{C}$ as being the {\em skeleton} of $\mc{C}$; and we
shall refer to the better understood set of special configurations $(\alpha, \beta_1, \beta_2)$,
with $\alpha$ null, $\beta_1$, and $\beta_2$ real, and $\alpha=\beta_1+\beta_2$, as being the universal
{\em scaffold} at the basis of our construction. 

As the name `skeleton' suggests, there are more constraints than those
associated to null roots. Below, we shall give explicit examples of (`fleshy') constraints
associated with strictly imaginary roots $\alpha^2 < 0$. However, constraints
associated to null roots play a distinguished role in our construction.
 The special role of light-like $\alpha$ is already suggested by the affine Sugawara 
 construction (\ref{sugaff2}) where constraints were {\it only} defined for null roots. 
 In addition, the special configurations where both $\beta_1$ and $\beta_2$ are real 
 introduce a significant simplification in our construction. Indeed, in that case
 the root spaces associated to $\beta_1$ and $\beta_2$ are one-dimensional, so that
 there exists a unique (up to sign) contraction between the associated step operators.
By contrast, when not both of $\beta_1$ and $\beta_2$ are real, the root spaces that are paired are multidimensional, and moreover not necessarily of equal dimension. This leaves
open many possibilities for `contracting' $T^{(s_1)}_{\beta_1}$ with $ T^{(s_2)}_{\beta_2}$
in forming $\mf{L}_{\alpha, \bar{s}}$. 
The information on how to contract the elements of different root spaces 
is then encoded in the choice of the coefficients $M_{s_1,s_2}(\beta_1,\beta_2)$.
Let us note, however, that, given a certain pair $(\alpha,\bar{s}) \in \mc{C}$,
{\it i.e.}, given a certain Lie algebra generator $T_{\alpha}^{(\bar{s})}$,
 there exists (when $\alpha=\beta_1+\beta_2$)
 a distinguished way of contracting (a part of) the 
 $\beta_1$ root space $\mf{g_{\beta_1}}$
 with the  $\beta_2$ one $\mf{g_{\beta_2}}$. Indeed, if we denote 
 $\beta_1 = \alpha - \beta$, so that $\beta_2 = + \beta$, the adjoint action of $T_{\alpha}^{(\bar{s})}$,
 ${\rm ad}_{T_{\alpha}^{(\bar{s})}} x \equiv \lb T_{\alpha}^{(\bar{s})}, x   \rb$
 maps  $\mf{g_{-\beta}}$ onto (a part of)  $\mf{g_{\beta_1}}=  \mf{g_{\alpha -\beta}}$. We can then
 use the natural `dual' pairing between $\mf{g_{-\beta}}$ and  $\mf{g_{+\beta}}$ ({\it i.e.},
 between $\mf{g_{-\beta_2}}$ and  $\mf{g_{+\beta_2}}$) to write putative constraints of the 
 form\footnote{To see that expression (\ref{sugadjoint}) is indeed well-defined, one can 
 invoke the invariance of the bilinear form, see Lemma~2.4 in~\cite{Kac}.}
 \be\label{sugadjoint}
\mf{L}_{\alpha, \bar{s}} = \sum _{\beta\in\Delta^{\text{hyp}}} \sum_{s}
N(\alpha,\beta)
\lb T^{(\bar{s})}_{\alpha}, T^{(s)}_{-\beta}\rb   T^{(s)}_{\beta} \,.
\ee
Here the coefficients $N(\alpha,\beta)$ no longer depend on the degeneracy index $s$
within the dual spaces $\mf{g_{\pm \beta}}$, and the sum over $s$ is easily seen to
be {\it independent} of the choice of (dual) bases $T^{(s)}_{\pm \beta}$ (as long
as the orthonormalization condition (\ref{hypcr}) is satisfied). We leave to future
work further study of the usefulness of the special construction (\ref{sugadjoint}).

One advantage  of expressing the constraints as in (\ref{sughyp}) is 
that, contrary to the expressions derived in~\cite{Damour:2007dt} (which were formulated
in terms of the $GL(10)$ level decomposition of $E_{10}$),
 such a definition a priori appears not to be tied to any particular level decomposition of the hyperbolic algebra. Therefore, this opens up the possibility of writing a `universal' set of
coset constraints, whose further (particular) level decompositions  could give rise to the 
apparently {\em different} canonical constraints arising in different maximal supergravities
(mIIA, IIB, $\ldots$). 
However, we shall give evidence below that this hope of a universal constraint construction
is not fulfilled in this simple way. Rather, we will encounter a more refined construction, where only the scaffold is universal. The reason appears to lie in the existence of various
ways of contracting (multi-dimensional) root spaces, {\it i.e.}, in the possibility of various
consistent choices for the coefficients $M_{s_1,s_2}(\beta_1,\beta_2)$. Each particular
level decomposition might be tied to a particular corresponding choice for these
coefficients. Even if this turns out to be the case, it seems that our construction
still involves a universal part, namely the part of (\ref{sughyp}) involving
the {\em skeleton} of `null' constraints, and its associated {\em scaffold} 
of special configurations where a null root $\alpha$ is decomposed into two real roots $\beta_1$ and $\beta_2$. As we shall emphasize below, this universal part 
is invariant under the Weyl group of the hyperbolic algebra and already yields
an infinite number of constraints (associated to the intersection of the light-cone 
with the root lattice). This `universal part' is, however, {\em not invariant} under the hyperbolic algebra itself. As we shall see below, one can associate to each choice of a finite-dimensional subalgebra (used as a way of `slicing' the hyperbolic algebra by means of a corresponding level decomposition) a way of generating additional constraints by covariantizing under that subalgebra. Each such covariantization procedure allows one to `flesh out' the skeleton by adding new constraints inside the light cone and also terms with $\beta_1$ and $\beta_2$ not both real. The prescription will be made more precise in section~\ref{Univsec} when we discuss the example of $E_{10}$.

A further general issue regarding  (\ref{sughyp}) is the operator ordering. Below we will work with similar 
expressions involving functions on classical phase space which are commuting. [Note
that they commute as functions, but do not `Poisson commute'.]
For those the issue of ordering becomes relevant only after the 
transition to the quantum theory, which we will not consider here. Finally, as 
written, (\ref{sughyp}) is meant to define only one constraint per root
even though null roots have multiplicity greater than one.

The structure of null roots in hyperbolic over-extended algebras is known to be given 
by Weyl orbits through
\be\label{hypnullrts}
\Delta^{\text{null}} = \bigcup_{n\in\ints\backslash\{0\}} \mc{W}\cdot (n \, \delta)\,,
\ee
where $\mc{W}$ is the hyperbolic Weyl group and $\delta$ the primitive 
null root of the affine algebra embedded in the hyperbolic extension. Restricting the construction (\ref{sugcharges}) to affine 
generators reduces all the Weyl orbits to points since $\delta$ is invariant 
under the affine Weyl group. Hence the construction 
gives constraints only for the roots $\alpha= n \, \delta$ in agreement with the 
affine Sugawara construction (\ref{sugaff2}).

At this point, we stress a possible qualitative difference between the usual affine Sugawara construction (\ref{sugaff2}) and the corresponding hyperbolic construction (\ref{sughyp}) at the
present stage of our understanding of the construction. 
The affine Virasoro constraints $L_{n\,\delta}$ form a
{\em two-sided} tower, where $n$ runs over the set of  integers
${\mathbb Z}$, while it seems consistent that the hyperbolic constraints $\mf{L}_{\alpha, \bar{s}}$
run over a set $\mc{C}$ which is a {\em one-sided} convex cone, contained within
the {\em past light-cone} of the Lorentzian root lattice. This one-sided
structure of the constraints was clearly apparent in~\cite{Damour:2007dt}, where only constraints $\mf{L}_\alpha$ corresponding to negative imaginary $\alpha$ were found, as will be shown in section~\ref{Univsec} below.\footnote{There was a further one-sidedness in~\cite{Damour:2007dt} related to the fact that we were working in a {\em truncated} coset whence only a Borel subalgebra of the hyperbolic algebra played a role. This effect is an artefact of the truncation and irrelevant to the present construction.}

 This asymmetry between the 
two-sidedness of the usual affine (Virasoro) constraints, and the one-sidedness 
of the hyperbolic ones,
seems to be deeply rooted in the different physics (and mathematics) associated
to the origin of these constraints. In the usual affine case, the origin of the
constraints is a gauge invariance under reparametrizations of (two) periodic
(world-sheet light-cone) variables $\sigma_{\pm} = \tau \pm \sigma$. The periodic
nature of these variables, and the real (or hermitian) character of the worldsheet
embedding functions, e.g. $\partial_{\pm} X^{\mu}(\tau,\sigma)$, implies the
existence of two-sided Fourier expansions involving, for each choice of sign
in $\sigma_{\pm}$ the two complex-conjugated
basis functions $\exp ( + i n \sigma_{\pm})$ and $\exp ( - i n \sigma_{\pm})$.
By contrast, the hyperbolic coset models should describe the gravitational
physics taking place near a spacelike singularity, {\it i.e.}, in a time-asymmetric
situation of the type  $t \to 0^+$, say. 
Moreover, the hyperbolic coset model is itself parametrized asymmetrically in terms of 
{\em positive} roots only. The analysis of the dynamics of supergravity in 
\cite{Damour:2002cu} found evidence for relating the supergravity fields
to {\em one-sided} towers of coset variables. This tower consists of the so-called `gradient generators' that are 
conjectured to correspond to multiple spatial gradients, roughly in terms of a spatial Taylor expansion. 
It is then  natural to conjecture that the usual space-dependent supergravity constraints will also give rise to one-sided-only
towers of `gradient cousins' of the (already one-sided) low-level  
constraints discussed in \cite{Damour:2007dt}. 

Another (related) argument for expecting that the tower of coset constraints 
be one-sided only, is the idea proposed in \cite{Damour:2007dt} that the set
of constraints be just large enough to reduce the exponentially infinite number
of variables entering the hyperbolic coset models to a much smaller number of
degrees of freedom involving only a rather small vicinity of the future light-cone
in root space ({\it i.e.}, essentially the gradient generators, plus a relatively manageable
set of extra M-theoretic degrees of freedom). To achieve such a strong reduction in the
number of degrees of freedom, without killing them all, it is natural to have a set
of constraints $\mc{C}$ which fills, like the coset variables, a one-sided
cone and whose degeneracies do not grow faster than the ones of the roots.
Note, however, that our intuitive argument cannot exclude the possibility that the
constraints fill a double-sided cone, if the degeneracies of the constraints
are such that the sum of the positive-sided and negative-sided ones does not
grow faster than the positive-root degeneracies.

Whatever be the ultimate definition of the physically correct set of
coset constraints, $\mf{L}_{\alpha, \bar{s}}$, one would expect it to
satisfy some commutation relations (of the general type 
$ \lb \mf{L}, \mf{L} \rb = O( \mf{L})$)
reflecting some aspects of the (currently unknown)
underlying gauge symmetry of the hyperbolic models, in the same way
that the Virasoro  algebra (\ref{affvir}) is a gauge-fixed remnant
of the worldsheet diffeomorphism symmetry of the underlying (Nambu-Goto-type)
string action.
Given the trial expression (\ref{sughyp}) one can wonder what algebra these 
expressions satisfy, {\it i.e.}, whether there is a generalization of the Virasoro 
algebra (\ref{affvir}) associated with our construction. While a conclusive 
answer to this question would require a knowledge of the $E_{10}$ algebra 
which is presently not available, we can at least formulate the following 
expectation. Under the Poisson (or Dirac) bracket the grading of the algebra 
implies that the simplest type of commutation relation one might have is of the form
\be\label{virhyp}
\left\{ \mf{L}_{\alpha} , \mf{L}_{\beta} \right\} = \sum_{\gamma} J_{\alpha+\beta-\gamma} \mf{L}_\gamma\,.
\ee
As we shall discuss in the next section below, relations of the type (\ref{virhyp}) {\em do hold}
if we consider only the (truncated, low-level) constraints of \cite{Damour:2007dt}.
However, the vast generalization of the definition of the constraints
introduced in the present paper makes the validity of a result of the type (\ref{virhyp})
highly non-trivial and dependent upon delicate structures that we do not
currently understand in detail. Indeed,
there are two non-trivial assertions contained in the expected result (\ref{virhyp}). 
The first one is that the trilinear\footnote{The hyperbolic Lie algebra
structure $\left\{ J, J \right\}=J$ guarantees that the commutator of two
$J$-bilinear constraints $\mf{L}$ is only trilinear in the $J$'s.} 
expression in current components on the 
r.h.s. organizes itself into products between constraints and certain
current components, much in the same way as for the affine 
Virasoro algebra (cf.~(\ref{affvir}) where the r.h.s. is a product of a
constraint $L_{n \delta}$ by a (conserved) algebra generator $c$).
The second claim relates 
to the roots $\gamma$ contributing on the r.h.s. and the question whether
these only cover constraints that had been defined previously. Both points 
are important for ascertaining the closure of the constraint algebra. The 
fact that only strongly conserved coefficients appear in the algebra of 
constraints is important for the discussion of open algebras, as mentioned 
in the introduction. We note one point concerning (\ref{virhyp}) in comparison to the affine Virasoro algebra (\ref{affvir}). There it was important that an additive structure existed on the set of all roots for which generators $L_{m\delta}$ were defined. Here, we expect that this additive
structure will be replaced by a certain convexity-related structure of the cone $\mc{C}$,
akin to the structure of integrable highest-weight representations \cite{Kac}. Though we do not
yet fully comprehend this structure, we shall see below that our proposed
`fleshing out' of the skeleton ensures (when starting from a past-light-cone-only
skeleton) the convex structure of a solid cone, {\it i.e.}, all $\alpha$'s  generated by our construction lie on or inside the light-cone.

\end{subsection}

\end{section}

\begin{section}{Universality and relation to supergravity}
\label{Univsec}

In this section we specialize to the case of $E_{10}$ whose Dynkin diagram is given in figure~\ref{e10dynk}. The relation to supergravity will help to make the construction of the preceding section more concrete. An important role will be seen to be played by the relation between $D=11$ supergravity (or type IIA in $D=10$),
 and type IIB in $D=10$.

\begin{subsection}{Consistency with supergravity constraints: $D=11$}

\begin{figure}[t]
\begin{center}
\scalebox{1}{
\begin{picture}(340,60)
\put(5,-5){$1$} \put(45,-5){$2$} \put(85,-5){$3$}
\put(125,-5){$4$} \put(165,-5){$5$} \put(205,-5){$6$}
\put(245,-5){$7$} \put(285,-5){$8$} \put(325,-5){$9$}
\put(260,45){$10$} \thicklines
\multiput(10,10)(40,0){9}{\circle{10}}
\multiput(15,10)(40,0){8}{\line(1,0){30}}
\put(250,50){\circle{10}} \put(250,15){\line(0,1){30}}
\end{picture}}
\caption{\label{e10dynk}\sl Dynkin diagram of $E_{10}$ with numbering
of nodes.}
\end{center}
\end{figure}
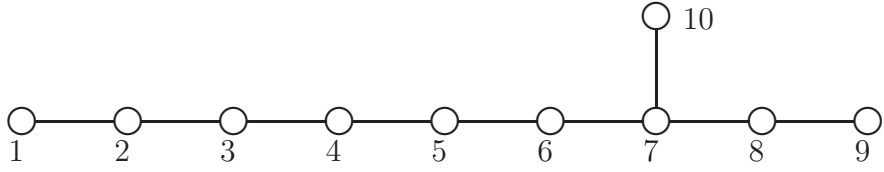

The Sugawara constraints (\ref{sughyp}) can be interpreted as constraints 
to be imposed 
on geodesics on the infinite-dimensional coset space $E_{10}/K(E_{10})$ as 
follows~\cite{Damour:2007dt}. The global $E_{10}$ symmetry gives rise to 
conserved Noether charges $\mc{J}\in\text{Lie}(E_{10})$ that can be expanded 
in the orthonormal basis $\{ T_\alpha^{(s)} \,| \,\alpha\in
\Delta^{\text{hyp}}, s=1,...,{\rm mult}\,\alpha \}$ as
\be\label{curexp}
\mc{J} =  \sum_{\alpha\in\Delta^{\text{hyp}}} 
 \sum_{s=1}^{{\rm mult}\,\alpha} J_{\alpha}^{(s)} T_{\alpha}^{(s)} \,.
\ee
The pairing between charges and generators is as in~\cite{Damour:2007dt}:
\be
\mc{J}&=&\ldots + \frac1{3!}\J{-1}{m_1m_2m_3} F_{m_1m_2m_3} + \J{0}{m}{}_n K^n{}_m \nn\\
&&\quad+ \frac1{3!}\J{1}{}{}_{\!\! m_1m_2m_3}E^{m_1m_2m_3}+\ldots\,,
\ee
where we have for definiteness chosen the $\mf{gl}(10)$ level decomposition 
of $E_{10}$ that is reviewed in appendix~\ref{a9ld}. An important point 
to note here is that tensor generators and coefficients transform 
contragrediently. For instance, for the Chevalley-Serre
generators this translates into the following identification
\begin{eqnarray}
T_{\alpha_1} &=& K^1{}_2 \quad\quad\qquad \big[ \sim e_1 \big] \nn\\
J_{\alpha_1} &=& J^2{}_1 \quad\quad\, \big[ \sim f_1 = -\omega (e_1) \big]
\end{eqnarray}
and so on, where $\omega$ is the Chevalley involution on $E_{10}$.
With this identification of algebra generators 
and current components we can work either in the universal enveloping 
algebra, generated by the $T_\alpha$, or in the Poisson algebra, generated 
by the current components $J_{\alpha}$. Namely, when considered as 
elements of a Poisson algebra on phase space, the components $J_\alpha^{(s)}$ 
close into the same hyperbolic algebra under Poisson commutation,
as follows directly from the Hamiltonian formulation of the coset 
space dynamics. That is, we have the canonical brackets
\be
\left\{ J_{\alpha_1}^{(s_1)}, J_{\alpha_2}^{(s_2)} \right\} = f_{\,\,\,\alpha_1\,\,\,\alpha_2\,\,\, \,(s_{12})}^{(s_1)(s_2)\,\alpha_1+\alpha_2} J_{\alpha_1+\alpha_2}^{(s_{12})}\,,
\ee
identical (including the sign) to the commutation relations of the hyperbolic algebra (\ref{hypcr}).
The classically conserved charges of the $E_{10}/K(E_{10})$ model are 
commuting functions on phase space in terms of which we write the 
classical constraints as
\be\label{sugcharges}
\mf{L}_\alpha = \sum_{\beta\in\Delta^{\text{hyp}}} 
\sum_{s,s'} M_{s,s'}(\alpha,\beta) J_{\alpha-\beta}^{(s)} J_{\beta}^{(s')}\,.
\ee
without specifying the summation over the `internal' degrees of freedom at
this point (that is, the matrix $M_{s,s'}(\alpha,\beta)$). 
The Hamiltonian (scalar) constraint entering the coset model 
of~\cite{Damour:2002cu} can be represented as the special member of the 
hierarchy of constraints (\ref{sugcharges}) corresponding to $\alpha =0$
\be\label{hcsug}
\mf{L}_0\equiv \mc{H} = \sum_{\beta\geq 0} 
    \sum_{s=1}^{{\rm mult}\, \beta} J_{-\beta}^{(s)} J_{\beta}^{(s)}\,.
\ee
In this way one confirms that all Noether charges $J_\alpha^{(s)}$ are 
indeed classically conserved because they Poisson commute with $\mc{H}$:
\be
 \left\{ \mc{H}, J_\alpha^{(s)} \right\} = 0\,.
 \ee
This is a direct consequence of the fact that  $\mc{H}$ is just the 
quadratic Casimir operator for the hyperbolic algebra (see chapter~2 of 
\cite{Kac} for a proof and the explicit computation). We note also that 
for the Hamiltonian constraint (\ref{hcsug}) the issues of contracting 
generators from root spaces of different dimensions are absent since 
the root spaces of $\alpha$ and $-\alpha$ always have the same dimension.
Since all components of $\mc{J}$ are conserved, {\em any} expression of 
the type (\ref{sugcharges}) is strictly conserved for any geodesic. We 
can therefore consistently constrain the geodesic motion on the coset 
space by demanding that the initial conditions satisfy $\mf{L}_\alpha=0$. 

In~\cite{Damour:2007dt} we have shown (with the same truncation of higher 
order spatial gradients as in \cite{Damour:2002cu}) that the canonical 
constraints of $D=11$ supergravity can be successively rewritten in two
different (but related) forms. 
Our analysis used an $A_9=\mf{sl}(10)$ level 
decomposition of the $E_{10}$ algebra, corresponding to the removal 
of node $10$ in fig.~\ref{e10dynk}. The results of this level decomposition 
of~\cite{Damour:2002cu,Nicolai:2003fw} are reproduced in appendix~\ref{a9ld}. 
The explicit computation involved the determination of 
various numerical coefficients in the $E_{10}$ expressions that were originally fixed 
by requiring weak conservation of the constraint surface under the coset model equations of motion.
Comparison with the canonical $D=11$ supergravity constraints and use 
of the dictionary then showed precise agreement of these numerical coefficients,
thus extending the correspondence between the $E_{10}/K(E_{10})$
coset model and the (truncated) $D=11$ supergravity equations of 
motion to the full canonical formulation. 
In section~\ref{normsec}, we shall show that, remarkably, these specific
numerical coefficients found for the supergravity constraints in~\cite{Damour:2007dt} {\em coincide} with our proposed sum over canonically normalized current components (\ref{sugcharges}) when both $\beta$ and $\alpha-\beta$ are real and for {\em unit} coefficients $M_{s,s'}(\alpha-\beta,\beta)$. 
 In addition to this unearthing of a hidden simplicity in the
 definition of the constraints, another advantage of writing the 
constraints  in the form (\ref{sugcharges}) is that this will allow us to
evaluate them also for other level decompositions, and in this way 
to verify agreement with the canonical constraints of massive IIA and 
IIB supergravity as well. The agreement between the dynamical
(evolution) equations of these theories with the coset model 
equations in appropriate truncations had already been established 
in~\cite{Kleinschmidt:2004dy,Kleinschmidt:2004rg,Henneaux:2008nr}. Moreover, the form (\ref{sugcharges}) is directly 
amenable to an affine reduction, and brings out more clearly the 
analogy with the affine Sugawara construction.

 \begin{subsubsection}{On the roots associated to the supergravity constraints}
 \label{nullsec}
 
Let us first turn to the detailed consideration of the set of
roots, including their multiplicities, that are associated to
supergravity constraints.
In the case of $D=11$ supergravity, these constraints
 are, respectively, the diffeomorphism and Gauss constraints, and the 
Bianchi identities for the 4-form field strength and the Riemann 
tensor.\footnote{In a more conventional canonical analysis, one would 
  not interpret the Bianchi identities as proper constraints, as they are 
  not directly associated to gauge transformations, unlike the diffeomorphism 
  and Gauss constraints. In the present setting, however, they would
  correspond to generators of gauge transformations on the {\em dual}
  fields, {\it i.e.}, on the 7-form field and the `dual graviton'.}
The analysis of~\cite{Damour:2007dt} was based on a  $\mf{gl}(10)$ level 
decomposition truncated at level $\ell=3$, such that, when expressed in terms
of the conserved $E_{10}$ Noether current in this decomposition, 
the constraints take the form
\begin{subequations}\label{oldconstraints}
\be\label{olddiff}
\cL{-3}{n_1\ldots n_9} &=& 28 \J{-1}{[n_1n_2n_3}\J{-2}{n_4\ldots n_9]} + 3 \J{-3}{p|[n_1\ldots n_8}\J{0}{n_9]}{}_{p}\,,
\ee
\be
\cL{-4}{m_1\ldots m_{10}||n_1n_2} &=& \frac{21}{10}\! \J{-2}{n_1[m_1\ldots m_5}\!\!\J{-2}{m_6\ldots m_{10}]n_2} +\frac32\!\!\J{-3}{n_2|[m_1\ldots m_8}\!\J{-1}{m_9m_{10}]n_1}\nn\\
\label{oldgauss}
&&  \qquad   \quad - (n_1\leftrightarrow n_2)\,,
\ee
for the diffeomorphism and Gauss constraints and
\be
\label{oldbianchiF}
\cL{-5}{m_1\ldots m_{10}||n_1\ldots n_5} &=& 3\J{-2}{m_1m_2[n_1\ldots n_4}\J{-3}{n_5]|m_3\ldots m_{10}}\,,\\
\label{oldbianchiR}
\cL{-6}{m_1\ldots m_{10}||n_0|n_1\ldots n_7} &=& 9 \J{-3}{n_0|m_1\ldots m_8}\J{-3}{m_9|m_{10}n_1\ldots n_7}\,.
\ee
\end{subequations}
for the Bianchi identities. Here, we have changed the normalization of the 
charge $J^{(-3)}$ compared to~\cite{Damour:2007dt,Damour:2004zy} so that 
all highest weight states are uniformly normalized to unity (the usefulness
of this re-definition was already pointed out in footnote~19 of~\cite{Damour:2007dt}). 
Explicitly, the normalizations of the $E_{10}$ generators, in their $A_9$ decomposition, are
\begin{eqnarray}
\langle \J{0}{a}{}_b | \J{0}{c}{}_d\rangle &=& 
 \delta^a_d\delta^c_b - \delta^a_b\delta^c_d\; ,\quad
 \langle \J{-1}{a_1a_2a_3} | \J{1}{}{}_{\!\!b_1b_2b_3}\rangle = 
 3! \,\delta^{a_1a_2a_3}_{b_1b_2b_3}\,, \nn\\ 
&&
\langle \J{-2}{a_1\ldots a_6} | \J{2}{}{}_{\!\!b_1\ldots b_6}\rangle = 
    6! \, \delta^{a_1\ldots a_6}_{b_1\ldots b_6}\,,
\end{eqnarray}
By contrast, for the mixed symmetry field on level $|\ell|=3$ 
we shall take here a
normalization that differs from the one given in Eq.~(2.30) of \cite{Damour:2004zy}
by a factor 1/9, {\it viz.}
\be
\langle \J{-3}{a_0|a_1\ldots a_8} | \J{3}{}{}_{\!\!b_0|b_1\ldots b_8}\rangle = \frac{8\cdot 8!}{9} \left(\delta^{a_0}_{b_0}\delta^{a_1\ldots a_8}_{b_1\ldots b_8}- \delta^{a_0}_{[b_1}\delta^{a_1\ldots a_7\,a_8}_{b_2\ldots b_8]b_0}\right)\,.
\ee
This normalization is chosen
so that operators associated to real roots (two indices identical) have unit 
norm, like the highest weight
\be
\langle  \J{-3}{10|3\,4\,5\,6\,7\,8\,9\,10} | \J{3}{}{}_{\!\!10|3\,4\,5\,6\,7\,8\,9\,10}\rangle =1
\ee
whereas for operators associated to null roots (all indices different)
\be
\label{8/9}
\langle  \J{-3}{2|3\,4\,5\,6\,7\,8\,9\,10} | \J{3}{}{}_{\!\!2|3\,4\,5\,6\,7\,8\,9\,10}\rangle =\frac89\,.
\ee

In addition to these normalizations, we have used in (\ref{oldconstraints}) the same implicit antisymmetrization conventions as in~\cite{Damour:2007dt}. For instance, the expression in (\ref{oldbianchiF}), corresponding to a Bianchi constraint on the four-form field strength, is understood to be 
antisymmetrized (with weight one) over $m_1\ldots m_{10}$; furthermore the last relation 
(\ref{oldbianchiR}) is to be projected  onto a $(7,1)$ hook for the 
indices $n_1\ldots n_7$ and $n_0$. We note that for the constraints listed in 
(\ref{oldconstraints}) there are no ordering ambiguities in a possible 
transition to operator expressions in a quantum theory, except for $\mf{L}^{(-6)}$ in (\ref{oldbianchiR}), since all commutator 
terms vanish by Jacobi or Serre relations; for instance
\be\label{Serre}
\left[\J{-1}{[m_1m_2m_3} \,,\, \J{-2}{m_4\dots m_9]}\right] \,\propto \,
 \J{-3}{[m_1|m_2\dots m_9]} = 0
 \ee

Let us now exhibit the roots  underlying the diffeomorphism constraint 
(\ref{olddiff}). For this, we first consider its highest component, corresponding to the indices 
$2\,3\,4\,5\,6\,7\,8\,9\,10$. To identify the root $\alpha$ to which 
it belongs we must find  the eigenvalues under the ten Cartan generators 
of $E_{10}$. (Indeed, the `covariant' components, $\alpha_i \equiv \alpha(h_i)$ of
a root precisely encode the eigenvalues in $\lb h_i, e_\alpha \rb = \alpha(h_i) e_\alpha$.)
Since we are working with the current components $J$ we display 
the Cartan elements in this description. In the $\mf{gl}(10)$ basis the 
Cartan elements are 
\be
h_i &=& J^i{}_i - J^{i+1}{}_{i+1}\quad \quad(i=1,\ldots,9)\,,\nn\\
h_{10} &=& -\frac13\left(J^1{}_1+\ldots+J^7{}_7\right)+\frac23\left(J^8{}_8+J^9{}_9+J^{10}{}_{10}\right) \,.
\ee
Alternatively, one can do the calculation with Lie algebra elements, using the more familiar expressions of the Cartan generators $h_i$ in terms of Lie algebra generators recalled  
in appendix~\ref{a9ld}. (In that case, one notes that the constraint 
$\cLtxt{-3}{2\,3\,4\,5\,6\,7\,8\,9\,10}$ is associated with the contragredient Lie-algebra basis element $F_{2\,3\,4\,5\,6\,7\,8\,9\,10}$.) 
An easy calculation shows that the only non-zero eigenvalue
corresponds to $h_1$ (first node in figure~\ref{e10dynk}), 
and is equal to $+1$.
Hence, the list of `covariant' components  $\alpha_i \equiv \alpha(h_i)$, also known as 
`Dynkin labels', is  $[+1,0,0,0,0,0,0,0,0,0]$. 
This is equivalent to saying that the root associated to the highest component of the
diffeomorphism constraint is equal to the fundamental weight $\Lambda_1$ associated 
  to the simple root $\alpha_1$.\footnote{\label{hstfn} The fundamental weights $\Lambda_i$ are defined as dual to the simple roots $\alpha_j$ w.r.t. the Cartan inner product: $\langle \Lambda_i |\alpha_j\rangle = +\delta_{ij}$.
The fact that $\Lambda_1$, and the integrable highest-weight representation $L(\Lambda_1)$ 
built from it, is related to the tower of constraints 
was already discussed at some length in \cite{Damour:2007dt}. This relation does
not mean, however, that $\cLtxt{-3}{2\,3\,4\,5\,6\,7\,8\,9\,10}$
is a highest weight vector for the action of all the $E_{10}$ generators. Actually,
as was already shown in
\cite{Damour:2007dt}, and will be further discussed in section~\ref{Remsec}, it fails to be one.}
To explicitly write the root $\alpha = \Lambda_1$ associated to the highest diffeomorphism
constraint in terms of the simple roots, we must convert its Dynkin labels to root labels,
{\it i.e.}, pass from covariant indices to contravariant ones by using the inverse of the
Cartan matrix $A_{i j} = \langle h_i| h_j \rangle$. This leads to the
corresponding root $\alpha=-(\alpha_2+2\alpha_3+3\alpha_4+4\alpha_5+5\alpha_6+6\alpha_7+4\alpha_8+2\alpha_9+3\alpha_{10})\equiv -\delta$, where the (positive) root $\delta$ denotes the primitive null root of 
$E_9\subset E_{10}$. In particular, this shows that the root $\alpha = \Lambda_1$ associated to the highest component of the diffeomorphism constraint is a {\em negative null} root.\footnote{We note that the association of the
`null' ( or `cusp') fundamental weight $\Lambda_1$ to the diffeomorphism 
constraint is valid not only for maximal supergravity and $E_{10}$,  but also for other (super)gravity theories. For instance, for pure gravity in any spatial dimension $d$
the basic (diffeomorphism) constraint is always associated to roots of the
form $-\mu_a$, where $\mu_a$ (with $a=1,\ldots,d$) denotes the null roots
that are contained within the $GL(d)$ multiplet of the `gravity root'.
The notation $\mu_a = - \beta^a + \sum_c  \beta^c$ is the notation used
in \cite{Damour:2002et}. Note that the null root
$- \mu_1$ is indeed the fundamental weight associated with
the `hyperbolic' node of $AE_d$ (as explicitly dispayed in equation (3.14)
of \cite{Damour:2001sa}).}
 We can
therefore write for this particular component
\be\label{Ldelta}
\cL{-3}{2\,3\,4\,5\,6\,7\,8\,9\,10}  \equiv  \mf{L}_\alpha    \qquad 
\mbox{with $\alpha =\Lambda_1=- \delta \equiv -\delta^{(3)}$}\,.
\ee

Let us now consider  consider the roots associated to the
other components of the diffeomorphism constraint (\ref{olddiff}). They are 
obtained by the action of the permutation group $\mc{S}_{10}$ on the 
indices. Since the permutation group is the Weyl group of $\mf{sl}(10)$, 
we conclude that {\em all} components of the diffeomorphism constraint 
are associated with (negative) null roots, forming a single orbit of the 
Weyl group $W(\mf{sl}(10))$. These null roots can be obtained by acting 
with the corresponding Weyl transformation on $\delta$, such that
\be
w\left(\mf{L}_\alpha\right) = \mf{L}_{w(\alpha)}
\ee
where $w$ on the left hand side acts on the indices of the constraint 
$\mf{L}$ by permuting them. 

Let us now proceed to considering the roots associated to the higher-level
(or rather `lower-level', as the levels are negative) constraints.
To find the roots for the level $\ell=-4$ and $\ell=-5$ constraints in (\ref{oldgauss}) and (\ref{oldbianchiF}), 
we consider their highest weight components. These are 
$\cLtxt{-4}{1\,2\,3\,4\,5\,6\,7\,8\,9\,10||9\,10}$ and 
$\cLtxt{-5}{1\,2\,3\,4\,5\,6\,7\,8\,9\,10||6\,7\,8\,9\,10}$, respectively. 
A straightforward calculation gives the eigenvalues 
$[0,0,0,0,0,0,0,1,0,-1]$ and $[0,0,0,0,1,0,0,0,0,-1]$, respectively.
The corresponding roots are again found to be {\em null} and negative.
In view of the fact, recalled in (\ref{hypnullrts}), that all null roots are Weyl images of the basic
one-dimensional string of affine null roots $n \, \delta$,
we can look for the specific affine root $n \, \delta$ from which they descend. 
We find that it is $ - \delta$, {\it i.e.}, $n=-1$. In other words, 
in addition to being null, the roots associated to the level $\ell=-4$ and $\ell=-5$ constraints
can be obtained from the `basic' $\ell=-3$ `diffeomorphism-constraint'
root $\alpha =\Lambda_1=- \delta \equiv -\delta^{(3)}$ by applying
some $E_{10}$ Weyl reflection:  $w_{\alpha}(\beta) = \beta- (\alpha\cdot\beta)\alpha$
(here simplified by taking into account the fact that $\alpha \cdot \alpha =2$
for the roots of a simply laced algebra). More explicitly, we have:
\be\label{delta4}
\delta^{(4)} = w_\theta (\delta^{(3)}) \,, \quad \theta 
:= \alpha_1 + \alpha_2 + \alpha_3 + \alpha_4
     + \alpha_5 + \alpha_6 +\alpha_7 + \alpha_{10}
\ee
and 
\be\label{delta5}
\delta^{(5)} = w_{\theta'}(\delta^{(4)})\,, \quad \theta':= \alpha_6 + 2\alpha_7 + 2\alpha_8 +
     \alpha_9 + \alpha_{10}
 \ee
where we have given the explicit Weyl reflections in $\mc{W}(E_{10})$ 
that move between the different levels. Note that $\theta$ is the highest 
root of the embedded $A_8$ algebra associated with the IIB theory,  
and $\theta'$ is the highest weight of an embedded $D_5$ algebra.
Finally, similarly to the case of the roots associated to $\mf{L}^{(-3)}$,
the fact that the Young tableaux describing the $GL(10)$ index structure
of  $\mf{L}^{(-4)}$ and  $\mf{L}^{(-5)}$ are totally antisymmetric
guarantees that {\em all} the roots associated to the other
components of these constraints are obtained from the basic ones (\ref{delta4})
and (\ref{delta5}) by $GL(10)$ permutations, {\it i.e.}, by further Weyl reflections.
In particular, all of them are {\em null}.

So far all the roots associated to the first three levels of constraints
have been found to be light-like (and negative).
The constraint $\mf{L}^{(-6)}$ differs from the lower level ones in
that it is the first in the hierarchy of constraints to involve a non-trivial 
Young tableau. As a consequence, we are going to see that it contains
a mixture of null ($\alpha^2=0$) and time-like ($\alpha^2=-2$) roots.
More precisely, the highest weight component
$\cLtxt{-6}{1\,2\,3\,4\,5\,6\,7\,8\,9\,10||10|4\,5\,6\,7\,8\,9\,10}$
is easily checked to be associated to a null root, which can be obtained
from $-\delta^{(5)}$ by the following Weyl transformation
\be
\delta^{(6)} = w_{\theta''}(\delta^{(5)})\,,\quad \theta''=\alpha_4+2\alpha_5+2\alpha_6+2\alpha_7+\alpha_8+\alpha_{10}
\ee
Here, $\theta''$ is the highest root of an embedded $D_6$ algebra. 
Covariantizing this 
component under the action of the $\mf{sl}(10)=A_9$ subalgebra gives a 
representation of $(7,1)$ hook type which is not a pure antisymmetric 
tensor unlike the constraints on levels $-3$, $-4$ and $-5$. 
{}From the point of view of the permutation group 
$\mc{S}_{10}=\mc{W}(\mf{sl}(10))$ this means that there are two separate 
orbits under $\mc{W}(\mf{sl}(10))$. The `outer' orbit consists 
of permutations of the lowest weight indices and corresponds to 
null roots of $E_{10}$. The inner orbit corresponds to imaginary $E_{10}$ 
roots with $\alpha^2=-2$. In terms of the supergravity constraint 
(\ref{oldbianchiR}) these two orbits correspond to cases when there 
are two identical indices on the $(7,1)$ hook part or when they are 
all different, respectively.
The `skeleton' of null roots $-\delta^{(3)}, -\delta^{(4)},-\delta^{(5)},\ldots$,
together with their multiples (discussed below) and their time-like descendants,
is sketched in figure~\ref{fig:skeleton}.

\begin{figure}[h]
\centering
\includegraphics[width=65mm]{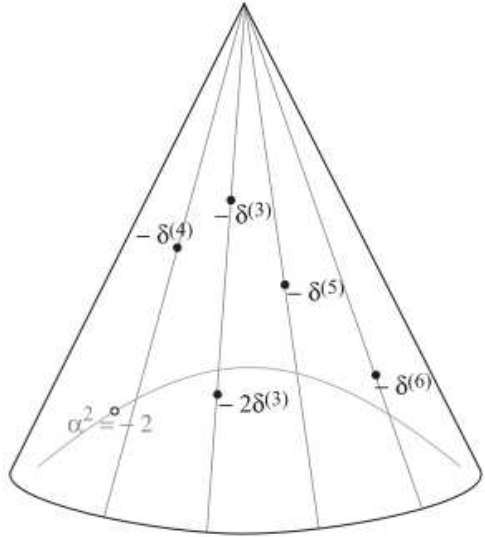}
\caption{\label{fig:skeleton}\sl Sketch of the set $\mc{C}$ of roots (and notably its `skeleton'
of null roots on the past light-cone) labelling the extended set
of constraints constructed in this paper.}
\end{figure}

Let us finally note that all null roots $\alpha$ appearing in these constraints appear with multiplicity one, although the same roots, considered as $E_{10}$ roots have the non-trivial root multiplicity eight. That the null roots appear with multiplicity one in the Sugawara construction should be so by consistency with the affine case. By contrast, the purely imaginary roots belonging to the inner orbit of $\mf{L}^{(-6)}$ have multiplicity seven as constraints compared to multiplicity $44$ as roots of $E_{10}$. 

\end{subsubsection}

\begin{subsubsection}{Supergravity constraints and canonical normalization}
\label{normsec}

So far we have analyzed the roots $\alpha$ labelling the l.h.s. of our basic
Sugawara-like expression (\ref{sughyp}).
Next we analyze the roots $\beta_1, \beta_2$ contributing to the {\em right hand side} of 
(\ref{sughyp}). Our principal aim here will be to see what are the
values of the numerical coefficients $M_{s_1,s_2}(\beta_1,\beta_2)$ that
enter the Sugawara-like sum.
We start here from the explicit $GL(10)$-decomposed form 
(\ref{oldconstraints}). To this aim let us consider the components 
of the currents $J$ on the r.h.s. where the indices 
are distributed in a specific way. For example,  we can pick out two 
representative terms where only operators for {\em real roots} appear and 
obtain
\be\label{diffcomp}
\cL{-3}{2\,3\,4\,5\,6\,7\,8\,9\,10}\!\! &\ni&\!\! 28\cdot\frac{3!\cdot 6!}{9!} 
\J{-1}{2\,3\,4}\J{-2}{5\,6\,7\,8\,9\,10} + 
3\cdot\frac{8!}{9!}\J{-3}{2|2\,3\,4\,5\,6\,7\,8\,9}\J{0}{10}{}_{2}\nn\\
&=& \frac{1}{3}\left( \J{-1}{2\,3\,4}\J{-2}{5\,6\,7\,8\,9\,10} + \J{-3}{2|2\,3\,4\,5\,6\,7\,8\,9}\J{0}{10}{}_{2}\right)\,.
\ee
Hence, we find the remarkable fact that the combinatorial factors 
appearing in (\ref{olddiff}) are precisely such 
as to imply, in the root basis, a relative normalization equal to {\em unity}. 
As the overall prefactor $1/3$ (as well as the corresponding  $1/60$ in the formulas below)
is merely chosen to agree with the normalisations in \cite{Damour:2007dt}, 
it might eventually be traded for a more convenient one.
Thus, all terms in the bracket 
belong to real roots and are canonically normalized, justifying {\em in retrospect}
the relative factor in (\ref{olddiff}) by (\ref{sugcharges}).

For the Gauss constraint (\ref{oldgauss}) one similarly finds
\be\label{gausscomp}
&&\cL{-4}{1\,2\,3\,4\,5\,6\,7\,8\,9\,10||9\,10} \,\ni\,
 \frac{21}{10}\cdot\frac{2\cdot5!\cdot 5!}{10!} \J{-2}{9\,1\,2\,3\,10\,4}\J{-2}{5\,6\,7\,8\,9\,10}\nn\\
&&\quad\quad + \,\frac{3}{2}\cdot\frac12\cdot\frac{2\cdot 8!}{10!}\J{-3}{9|9\,10\,1\,2\,3\,4\,5\,6}\J{-1}{7\,8\,10}\\
&&\quad= \frac1{60}\left(\J{-2}{9\,1\,2\,3\,10\,4}\J{-2}{5\,6\,7\,8\,9\,10}+\J{-3}{9|9\,10\,1\,2\,3\,4\,5\,6}\J{-1}{7\,8\,10}\right)\,.\nn
\ee
Again, the terms appear with the same relative coefficient and confirm the expression (\ref{sugcharges}) for real roots. 

For the constraints (\ref{oldbianchiF}) and (\ref{oldbianchiR}) on levels 
$\ell=-5$ and $\ell=-6$ there is nothing to check since there is only one 
type of term. The basic `scaffold' of Sugawara constraints exhibiting a 
decomposition $\alpha=\beta_1+\beta_2$ with $\alpha$ null and 
$\beta_1, \beta_2$ real is illustrated in figure~\ref{fig:scaffold}.
Note that the relations $\alpha^2=0$ and $\beta_1^2=\beta_2^2=2$ imply that $\alpha \cdot \beta_1 = 0 = \alpha \cdot \beta_2$,
{\it i.e.}, that $\beta_1$ and $ \beta_2$ are orthogonal to $\alpha$, so that they belong to the hyperplane
tangent to the light-cone along the considered null root  (see figure~\ref{fig:scaffold}, 
where one has chosen
$\alpha = - \delta$). One has to imagine the infinite `scaffold' made by the
tangent hyperplanes associated to 
the infinite skeleton of Weyl images of $- \delta$.

\begin{figure}[h]
\centering
\includegraphics[width=65mm]{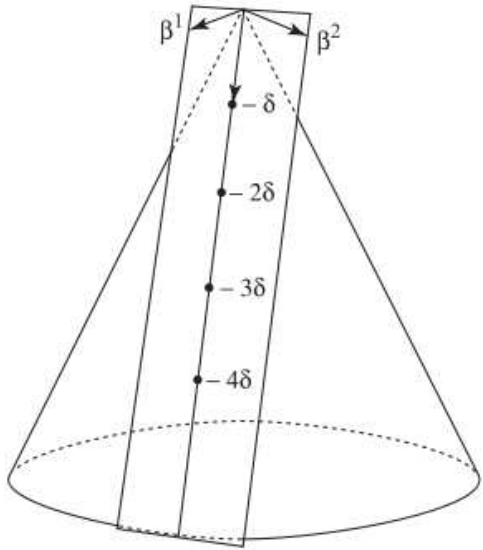}
\caption{\label{fig:scaffold}\sl Sketch of one of the basic elements of the infinite `scaffold' of 
special Sugawara configurations $\alpha=\beta_1+\beta_2$ with $\alpha$ null and $\beta_1,\beta_2$ real.
The real roots $\beta_1, \beta_2$ lie within the hyperplane tangent to the light-cone
along the null root (here chosen to be $\alpha = - \delta$). One must imagine completing
the structure shown here by all its Weyl images.}
\end{figure}

\end{subsubsection}

\begin{subsubsection}{General structure of constraints}

We note that there are also terms contributing to (\ref{sugcharges}) where not both of $J_{\alpha-\beta}$ and $J_\beta$ are real.  For example, (\ref{olddiff}) contains a term
\be
\cL{-3}{2\,3\,4\,5\,6\,7\,8\,9\,10} &\ni& \frac13 \J{-3}{1|2\,3\,4\,5\,6\,7\,8\,9}\J{0}{10}{}_1\,,
\ee
where an imaginary level three root is contracted with a real level zero root (albeit positive). Similar contractions appear also for the other constraints. 
Note that, though, after removing the same prefactor ($1/3$) as above, we have
again a simple coefficient unity, the time-like-root generator associated to
the level $-3$ root is such that its normalization involves the fraction $8/9$,
see (\ref{8/9}).

At this stage, we start seeing several patterns appearing within the
structure of the constraints, and notably in the set $\mc{C}$ labelling the
roots (together with their multiplicity) associated to the constraints.
A first pattern is that, so far, all the constraints can be labelled by the
members of the integrable highest-weight representation descending from the
fundamental weight $\Lambda_1$, which is dual to the first (`hyperbolic')
node of the Dynkin diagram, figure~\ref{e10dynk}.
A second, closely related, pattern is that the pattern of roots comprise
many null roots, and that the null constraint-roots studied so far all belong 
to the Weyl orbit of $\Lambda_1= -\delta$. 
A third pattern is the simple (unit) relative normalization of the contributions
$(\alpha,\beta_1,\beta_2)$ to the Sugawara expression (\ref{sughyp}) involving the decomposition
of a null root $\alpha$ into two real roots $(\beta_1,\beta_2)$.
A fourth pattern is that the null roots associated to non purely antisymmetric
Young tableaux give rise, upon covariantization under $GL(10)$, to a set
of roots which `penetrate' within the past light-cone, {\it i.e.}, which are
time-like (and past-directed) rather than light-like.

It is tantalizing to generalize these patterns to the infinite tower of
coset constraints that we are trying to construct.
We can first assume that the set $\mc{C}$ of `constraint roots' contains
the full weight diagram, say $P(\Lambda_1)$,
of the fundamental representation  $L(\Lambda_1)$ based on $\Lambda_1= -\delta$.
By proposition~10.1 of~\cite{Kac} we know that $P(\Lambda_1)$ (including
its multiplicities) is invariant under  the full $E_{10}$ Weyl group, $\mc{W}(E_{10})$.
In particular, this would imply, in view of (\ref{hypnullrts}), that 
 there is an {\em infinite 
sequence of null constraints} related to the orbit of minus the primitive null element $-\delta$.
Upon covariantization of the resulting highest weight vectors under $\mf{sl}(10)$ we 
obtain a series of constraints related to $\delta$ as indicated in the first row of 
the following table

\begin{center}
\begin{tabular}{|c|c|c|c|c|c|c|c|}
\hline&$\ell=-3$ & $\ell=-4$ & $\ell=-5$ & $\ell=-6$ & $\ell=-7$ & $\ell=-8$ &\ldots \\\hline\hline
$-\delta$ & $\mf{L}_{(-\delta)}^{(-3)}$ & $\mf{L}_{(-\delta)}^{(-4)}$ & $\mf{L}_{(-\delta)}^{(-5)}$ & $\mf{L}_{(-\delta)}^{(-6)}$ & $\mf{L}_{(-\delta)}^{(-7)}$& $\mf{L}_{(-\delta)}^{(-8)}$ &\ldots \\[2mm]
$-2\delta$ & &&& $\mf{L}_{(-2\delta)}^{(-6)}$ & &$\mf{L}_{(-2\delta)}^{(-8)}$ &\ldots \\
$\vdots$&&&&&&&\\\hline
\end{tabular}
\end{center}

Here, we added a subscript $(-\delta)$ to all the constraints in the $\mc{W}(E_{10})$ orbit 
of $-\delta$ and suppressed the labels for the $\mc{W}(\mf{sl}(10))$ suborbits in the columns.
Let us also recall the existence of the Hamiltonian constraint, $\mf{L}_0$, 
which could be thought of as being associated to the 0th multiple of   $\delta$.

In addition to the `skeleton' of null roots constituting the Weyl orbit of 
$\Lambda_1= -\delta$, the weight diagram  $P(\Lambda_1)$ of $L(\Lambda_1)$ contains {\em all
(past-directed) time-like roots}. This follows from proposition~11.2a of~\cite{Kac}. To apply this proposition, we need,
for each putative weight $\mu$ within the Weyl chamber ($ \mu \in P_+$; i.e. $\mu=\sum_{i=1}^{10} p_i \Lambda_i$, with $p_i\geq 0$),
to control the `support' of the root $\Lambda_1 - \mu$, i.e. the non-zero coefficients $m_j$ in its simple-root decomposition:
$\Lambda_1 - \mu = \sum_{j=1}^{10} m_j \alpha_j$. Using $\langle \Lambda_1 |\alpha_j \rangle=\delta_{1j}$, and $\langle \alpha_i | \alpha_j\rangle= A_{ij}$,
the root-basis integers $m_j$ are easily seen to be related to the weight-basis integers $p_i$ via the knowledge of the
{\em inverse} of the  $E_{10}$ Cartan matrix $A_{ij}$. Now, by explicit inspection of this inverse Cartan matrix (see,
e.g, \cite{KMW}), one finds
that the only place in it where there is a zero in the first column is in the first row. This shows
that  any element of the Weyl chamber $\mu=\sum_{i=1}^{10} p_i \Lambda_i$ such that $p_i\neq 0$ for at least one $i$ among
$2,\ldots,10$, the vector $\Lambda_1-\mu$ has non-vanishing `support' $m_1$
on the first node and hence is `non-degenerate w.r.t $\Lambda_1$' (in the sense defined in section 11.2 of~\cite{Kac}).
Hence, by Kac's proposition~11.2a such $\mu$'s are indeed weights (together with their Weyl images). The only exceptional
case is  when $p_j=0$ for $j=2,\ldots,10$, which corresponds to $\mu=p_1\Lambda_1$. In other words, we have found that all the negative
time-like weights belong to $P(\Lambda_1)$, but that the multiples of $\Lambda_1=-\delta$ are not part of the weight diagram $P(\Lambda_1)$.\footnote{Another way of seeing this is by using proposition~11.3 of~\cite{Kac} where  $P(\Lambda_1)$ is described as the convex hull of the Weyl orbit of $\Lambda_1$. The infinitely many Weyl images of $\Lambda_1$ all lie on the light-cone (and densely approximate any null direction)
and one might think that the convex hull covers all points on the light-cone. This is not true since one is constructing the convex hull as an infinite union of closed sets but this is not necessarily closed. In the present case it is open and misses exactly the multiples of $\Lambda_1$ and their Weyl images but the convex hull covers all points inside the light-cone.}

Though the set $P(\Lambda_1)$ is already quite large, it only corresponds to
the $GL(10)$ covariantization of the first row in the table above.
In view of the structure of the usual affine Virasoro-Sugawara constraints $L_{n \, \delta}$
recalled above, together with the known 
structure of $E_{10}$ null roots (\ref{hypnullrts}), it is now quite natural to
conjecture that the `null skeleton' of
$\mc{C}$ contains, in addition to the orbit of $- \delta$
(first row in the table) the Weyl orbits of (negative) multiples of $\delta$:  $-n\delta$.
This amounts to conjecturing that, besides the weight diagram $P(\Lambda_1)$ of the
fundamental representation $L(\Lambda_1)$, we must add the weight diagrams $P(n \Lambda_1)$
(with $n=2,3,\ldots$) corresponding to the multiple tensor product of $L(\Lambda_1)$
with itself:  $L(\Lambda_1)\otimes L(\Lambda_1)$,   $L(\Lambda_1)\otimes L(\Lambda_1)\otimes L(\Lambda_1)$, etc..

Besides this mathematical argument for conjecturing an extension of the set of
constraints beyond the ones related to the Weyl orbit of $- \delta$ (and its
covariantization), there is a physical argument suggesting the necessity of
this extension. Indeed, all the constraints discussed so far correspond, in view
of the `dictionary' of \cite{Damour:2002cu}, to the values {\em at one spatial point},
of some space-dependent supergravity constraints. For instance, 
$\cLtxt{-3}{n_1\ldots n_9}$ is the spatial $\epsilon^{n_1\ldots n_{10}}$ dual of the
diffeomorphism constraint $\mc{H}_m({\bf x}_0)$, taken at the specific spatial point ${\bf x}_0$
around which one analyzes the asymptotic behaviour of the supergravity fields
as $t \to 0$. However, the full supergravity diffeomorphism constraint
consists of imposing the vanishing of   $\mc{H}_m({\bf x})$ at all spatial points.
When expanding the diffeomorphism constraint  $\mc{H}_m({\bf x})$
in a (ten-dimensional) spatial Taylor expansion around the base point ${\bf x}_0$,
we see that we should replace the unique constraint $\mc{H}_m({\bf x}_0) \sim \cLtxt{-3}{n_1\ldots n_9}$
by an {\em infinite gradient tower} of spatial derivatives of the form
 $\partial_{m_1 \ldots m_k} \mc{H}_m({\bf x}_0)$. For instance, at the first spatial-gradient level
 $m=1$, we should be considering the two irreducible $GL(10)$ tensors 
 contained in $\partial_m \mc{H}_n({\bf x}_0)$, {\it i.e.}, its symmetric and antisymmetric parts. 
 Dualizing back these first-gradient constraints by means of
 $\epsilon^{n_1\ldots n_{10}}$, we are led to expecting that the
 `first-gradient descendants' of $\cLtxt{-3}{n_1\ldots n_9}$ will comprise
two  $GL(10)$ tensors bearing 18   contravariant indices, and belonging to two
 different Young tableaux: one with [9,9] boxes (corresponding to the symmetric combination) and one with [10,8] boxes (corresponding to the antisymmetric combination). 
 The former corresponds to the null root $-2\delta=2\Lambda_1$, whereas the latter corresponds to the imaginary $\Lambda_2$ and so lies inside the past light-cone.
  
 The extension of this
 gradient construction to the other supergravity constraints (Gauss, etc.)
  then naturally leads us to conjecture the existence of
 the second row of the table. Then, when considering higher spatial gradients
 we are led to conjecturing the existence of further rows `stemming' from
 $-3 \, \delta$, $-4 \, \delta$, etc. One finds that the putative 
constraints associated with the Weyl orbit of $-n\delta$ start on level $\ell=-3n$ and 
are spaced by $n$. Finally, it seems that the full table is describing all possible weights on or inside the (past) light-cone.

The notation in the table is condensed and does not display the 
$\mf{sl}(10)$ representation structure of the various constraints. For example, the set of
constraints labelled by $\mf{L}^{(-6)}_{(-\delta)}$ and $\mf{L}^{(-6)}_{(-2\delta)}$ transform 
in different $\mf{sl}(10)$ representations. The former one is in the hook representation of (\ref{oldbianchiR}), 
whereas the latter 
has  two sets of antisymmetric 9-tuples.  
Explicitly, one has the following two index structures 
\be\label{ellmoins6}
\stackrel{(-6)}{\mf{L}}{}_{\!\!\!(-\delta)}^{\! m_1\ldots m_{10}||n_0|n_1\ldots n_7}\, \text{and}
\stackrel{(-6)}{\mf{L}}{}_{\!\!\!(-2\delta)}^{\! m_1\ldots m_{9}|n_1\ldots n_9}\,.
\ee
In the affine truncation to $E_9$  only one member in each infinite sequence (row) 
for a given $-n\delta$ is non-trivial because of the presence of 10-tuples of
antisymmetrized indices in the higher components. In the example (\ref{ellmoins6}) above, the first tensor vanishes in the affine truncation, whereas the second one is non-zero. In addition, all the surviving constraints from the beginning of each sequence reduce to singlets under 
$\mf{sl}(9)$. These are the $\mf{L}^{(-3n)}_{(-n\delta)}$. This (one-sided)
sequence of constraints naturally correspond to the 
generators $L_{-n\delta}$ (for $n>0$) of the affine Sugawara construction that we had introduced in (\ref{sugaff2}). (We will return below to specific issues concerning 
the  contractions of the null roots and Cartan subalgebra generators).

We do not present an explicit expression for the second rung of constraints, like the second term in (\ref{ellmoins6}), but note that on the contractions of real root spaces it is given by the same general formula (\ref{sugcharges}) as the other constraints we have considered so far. 
 Among the other constraints in $\mf{L}^{(-n\ell)}_{(-n\delta)}$, some 
have an index structure similar to the elementary $\mf{L}^{(-\ell)}_{(-\delta)}$, but with all tuples  replicated $n$ times. For example, the index structure of $\mf{L}^{(-8)}_{(-2\delta)}$  contains a tensor with two $10$-tuples and two $2$-tuples
\be
\stackrel{(-8)}{\mf{L}}{}_{\!\!\!(-2\delta)}^{\!m_1\ldots m_{10}||p_1\ldots p_{10}|| n_1n_2|q_1q_2}\,.
\ee

To complete this discussion, let us point out the following `experimental' relation between  the constraints and the level 
decomposition of the adjoint of $E_{10}$ under $A_9$~\cite{Nicolai:2003fw}. 
`Admissible' $A_9$ representations in the level decomposition rarely appear 
with outer multiplicity zero. Here, `admissible' refers to solving necessary diophantine
conditions on the lowest weight vectors of a possible $A_9$ representation occurring in the adjoint representation of $E_{10}$, see eqns.~(6) and (7) in~\cite{Damour:2002cu}.
The only cases up 
to $\ell\leq 28$ for which the outer multiplicity of an admissible 
representation is zero are those when the associated lowest root in the 
representation is null.\footnote{This is no longer necessarily true when 
considering Kac--Moody algebras different from $E_{10}$~\cite{Kleinschmidt:2003mf} 
or decompositions other than that under $A_9$.} 
More precisely, the only entries with vanishing outer multiplicities in 
the tables of~\cite{Nicolai:2003fw} occur at\footnote{We use two different notations for describing elements $\alpha$ of the (self-dual) $E_{10}$ root lattice, namely in terms of either the basis of simple roots $\alpha_i$ or of the basis of fundamental weights $\Lambda_i$: $\alpha=\sum_i m_i \alpha_i = \sum_i p_i \Lambda_i$. In the former we write the ten-tuple of coefficients with round parentheses $(m_1,\ldots, m_{10})$ and in the latter with square brackets $[p_1,\ldots,p_{10}]$. The $p_i$ are often referred to as Dynkin labels. The $A_9$ weight is obtained from $[p_1,\ldots,p_{10}]$ by dropping the last entry $p_{10}$ since this corresponds to the node that is deleted in the $A_9$ level decomposition.}

\begin{center}
\begin{tabular}{|c|c|c|}
\hline
Level & $E_{10}$ root & $A_9$ weight\\
\hline\hline
$\ell = 3n$ & $n (0,1,2,3,4,5,6,4,2,3)$ & $[n,0,0,0,0,0,0,0,0]$\\
$\ell = 4n$ & $n (1,2,3,4,5,6,7,4,2,4)$ & $[0,0,0,0,0,0,0,n,0]$\\
$\ell = 5n$ & $n (1,2,3,4,5,7,9,6,3,5)$ & $[0,0,0,0,n,0,0,0,0]$\\\hline
\end{tabular}
\end{center}

The first line corresponds to the root $n\delta$ and both the second and the 
third line can be obtained from the first line by the Weyl transformations
given explicitly in eqs.~(\ref{delta4}) and (\ref{delta5}). These entries 
have vanishing outer multiplicities since the corresponding $E_{10}$ generators 
are already contained in the gradient representations on the relevant level. 
One potentially important implication of the vanishing outer multiplicities
is  that there are no ordering ambiguities because the relevant commutators
always vanish, as in (\ref{Serre}), whereas ordering ambiguities will occur 
in general for higher level constraints like $\mf{L}^{(-6)}_{(-\delta)}$.

\end{subsubsection}

\begin{subsubsection}{Algebra of constraints}

Let us now return to the question of the constraint algebra (\ref{virhyp}) raised at the end of section~\ref{sugcon}. We discuss this issue by using the 
explicit expressions for the constraints (\ref{oldconstraints}). As discussed above, one would like the constraint algebra to close with structure constants given by current components. From the results of~\cite{Damour:2007dt} it follows that one can generate higher level constraints from lower level constraints by the action of the {\em negative level}
current operator, $J^{(-1)}$, {\it i.e.}, that schematically
\be\label{poissonmodule}
\cL{-(\ell+1)}{} = \left\{\J{-1}{}, \cL{-\ell}{}\right\}\,,
\ee
is valid for $\ell=-3,-4,-5$. This property is equivalent to the result of~\cite{Damour:2007dt}
that the constraints are `covariant' under the upper Borel group $E_{10}^+$ ({\it i.e.}, that they
form a representation of $E_{10}^+$; even if they do not form a representation of the
full group $E_{10}$).
In addition, the {\em level-three truncated} constraints (\ref{oldconstraints}) have the
property that their Sugawara expression contains {\em only negative level} currents,
{\it i.e.}, schematically
\be\label{negativesuga}
\cL{-\ell}{} = \sum_{p+q=\ell} \J{-p}{} \cdot \J{-q}{} \,,
\ee
It is now easy to see that the two properties (\ref{poissonmodule}) and (\ref{negativesuga})
imply that the Poisson bracket of two constraints closes in the desired manner of (\ref{virhyp}).
This is certainly an encouraging result, which suggests that the structure of the
constraints incorporates special features allowing for the existence of a closed
algebra of the type of a generalized Virasoro algebra (\ref{virhyp}).

However, it is not clear whether the two special  properties (\ref{poissonmodule}) and (\ref{negativesuga}) continue to hold for the generalized infinite tower of
 $E_{10}$ constraints whose construction was sketched above. We shall see that the
 property (\ref{negativesuga}) is likely to be violated when implementing a
 certain `see-saw' construction defined below. As for the property 
 (\ref{poissonmodule}) (which says that the contraints form a representation of $E_{10}^+$),
 one reason for believing that it might not be universally valid  comes from the 
 example of the affine Sugawara construction. There the constraints do {\em not} transform in a representation of the affine algebra: $\mf{L}^{(-\ell-n)} \neq \left\{J^{(-n)}, \mf{L}^{(-\ell)} \right\}$.
  Rather one finds that it is the
  algebra which transforms under the constraints, {\it i.e.},
$J^{(-\ell-n)} = \left\{ J^{(-n)}, \mf{L}^{(-\ell)} \right\}$. 
We leave to future work further discussion of this important issue.

\end{subsubsection}

\end{subsection}

\begin{subsection}{Universality: $D=11$, IIB and massive IIA}

The full $E_{10}$ Lie algebra can be obtained from  the closure (via commutators)
of two of its finite-dimensional sub-algebras: (i) its $A_9$ subalgebra 
(relevant for $D=11$ supergravity), and (ii) its $A_8\oplus A_1$ subalgebra 
(relevant for type IIB supergravity). The $A_9$ subalgebra corresponds 
to nodes $1,2,3,4,5,6,7,8,9$ of the Dynkin diagram in fig.~\ref{e10dynk}; 
the $A_8\oplus A_1$ algebra corresponds to the nodes $1,2,3,4,5,6,7,10$ 
and $9$ in  fig.~\ref{e10dynk}. The two subalgebras $A_9$ and $A_8\oplus A_1$ 
together cover all ten nodes of the $E_{10}$ diagram, and therefore their 
closure is all of $E_{10}$. For the $A_8\oplus A_1$ decomposition, the term 
`level' refers to node $8$. For low levels, the decomposition under this 
$A_8\oplus A_1$ subalgebra, originally performed 
in~\cite{Kleinschmidt:2003mf,Kleinschmidt:2004rg}, is reproduced in 
appendix~\ref{a8a1ld}. 

The two decompositions under $A_9$ and $A_8\oplus A_1$ provide two different 
bases for the same Lie algebra $E_{10}$. In order to distinguish them 
we use the letter $J$ for the current components in the $A_9$ decomposition, 
as already done for example in (\ref{oldconstraints}), and the letter 
$I$ for current components in the $A_8\oplus A_1$ 
 decomposition. Since the real root spaces are one dimensional it is usually 
 straightforward to explicitly work out  the `change of basis' between the current 
 components expressed in the $J$ basis or the $I$ basis. For example, the root 
 space of the real root $\alpha=-\alpha_{10}$ contains the current component 
 $J^{8\,9\,10}$ in the $A_9$ decomposition. In the $A_8\oplus A_1$ decomposition 
 this root space is part of the $A_8$ `gravity line' and therefore one obtains the following 
 relation between the vectors of the two bases in the $\alpha=-\alpha_{10}$  root space
\be\label{basechangeex}
\J{-1}{8\,9\,10}  = \Jb{0}{8}{}_{9} \quad\text{corresponding to}\quad E_{-\alpha_{10}}\equiv f_{10} \,.
\ee
That the two generators are not on the same level with regard to the two
decompositions of the $E_{10}$ algebra will be of crucial importance for the 
construction we shall discuss next.
 In appendix~\ref{ldapp}, we also recall the association of the level 
decompositions with low-lying generators in an explicit tensor basis for the two 
decompositions.

The fact that $A_9$ and $A_8\oplus A_1$ together generate the whole $E_{10}$ algebra allows 
in principle to extend the lowest level supergravity constraints to arbitrarily high levels by the 
following mechanism (which for obvious reasons we will refer to as a {\em `see-saw mechanism'}).
Among the root components contributing to a given known constraint in one level decomposition, 
there are some that correspond to `unknown' levels in a different decomposition. 
Covariantizing the resulting expression with regard to the $\mf{gl}(n,\reals)$ subalgebra relevant for that 
new decomposition we generate new components, which in turn can be analyzed in terms 
of the first decomposition. Covariantizing again, but now with respect to the first decomposition,
we again generate new components. It is easy to see that this procedure never stops, and
so continues {\em ad infinitum}. 

To see how this construction works in a concrete example consider 
the following terms in the $D=11$ diffeomorphism constraint (\ref{olddiff}),
see also (\ref{diffcomp}),
\be\label{diff11part}
\cL{-3}{2\,3\,4\,5\,6\,7\,8\,9\,10} &\ni& 
\frac13\left(\J{-1}{2\,3\,4}\J{-2}{5\,6\,7\,8\,9\,10} \; + 
\J{-3}{9|2\,3\,4\,5\,6\,7\,9\,10}\J{0}{8}{}_9\right.\nn\\
&&\left.+\J{-3}{9|2\,3\,4\,5\,6\,7\,8\,9}\J{0}{10}{}_9 \;
-\J{-3}{8|2\,3\,4\,5\,6\,8\,9\,10}\J{0}{7}{}_8\right)\,.\quad\quad
\ee
All the terms in the bracket correspond to canonically normalized real root components 
of the current. In analogy with (\ref{basechangeex}) one can now convert these terms into 
the alternative basis provided by the $A_8\oplus A_1$ decomposition. In this way
we obtain (see appendix~\ref{ldapp} for the notation)
\begin{align}
\J{-1}{2\,3\,4} &= \Jb{-2}{2\,3\,4\,9}\,,&\quad \J{-2}{5\,6\,7\,8\,9\,10} &=\Jb{-2}{5\,6\,7\,8}\,,&\nn\\
\J{-3}{9|2\,3\,4\,5\,6\,7\,9\,10} &= \Jb{-3}{2\,3\,4\,5\,6\,7,\dot{1}}\,,&\quad \J{0}{8}{}_9 &=\Jb{-1}{8\,9,\dot{2}}\,,&\nn\\
\J{-3}{9|2\,3\,4\,5\,6\,7\,8\,9}&=\Jb{-4}{2\,3\,4\,5\,6\,7\,8\,9,\dot{1}\dot{1}}\,,& \quad \J{0}{10}{}_9&=\Jb{0}{\dot{2}}{}_{\dot{1}}\,,&\nn\\
\J{-3}{8|2\,3\,4\,5\,6\,8\,9\,10} &=\Jb{-4}{8|2\,3\,4\,5\,6\,8\,9}\,,&\quad \J{0}{7}{}_8&=\Jb{0}{7}{}_8\,.&
\end{align}
where dotted indices refer to the $\mf{sl}(2,\reals)$ algebra associated with node 9.
Putting this back into (\ref{diff11part}) one can see that this is part of a 
$GL(9,\reals)\times SL(2,\reals)$ covariant expression of the form\footnote{The 8-index
tensor on the l.h.s. is fully antisymmetric. We use the convention that $\epsilon^{\dot{1}\dot{2}}=+1=-\epsilon_{\dot{1}\dot{2}}$.}
\be\label{IIBdiff}
\Lb{-4}{n_1\ldots n_8} & =&\frac{35}{3}\Jb{-2}{[n_1\ldots n_4}\Jb{-2}{n_5\ldots n_8]}- \frac{28}{3}\Jb{-1}{[n_1n_2,\alpha}\Jb{-3}{n_3\ldots n_8],\beta} \epsilon_{\alpha\beta} \nn\\
&& -\frac13 \Jb{-4}{n_1\ldots n_8,\alpha\gamma}\Jb{0}{\beta}{}_\gamma\epsilon_{\alpha\beta}
- \frac{8}{3}\Jb{-4}{p|[n_1\ldots n_7}\Jb{0}{n_8]}{}_p +\ldots\,.
\ee
where for clarity of notation we use the symbol $\mf{C}$ to denote the 
IIB constraints. Remarkably, this expression is {\em exactly the diffeomorphism 
constraint of IIB supergravity} when the correspondence with $E_{10}$ 
of~\cite{Kleinschmidt:2004rg} is used. This is explained in more detail in appendix~\ref{IIBapp}.\footnote{We take this
  opportunity to point out a typo in the Einstein equation~(67) 
  in~\cite{Kleinschmidt:2004rg}: The terms involving the (self-dual) 
  five-form field strength should be multiplied by 1/2. This does not 
  affect the dictionary derived in that paper.} Indeed, using
the expressions (\ref{IIBCS1}) and (\ref{IIBCS2}) for the Cartan 
generators expressed in IIB variables, one finds that the component 
${2\,3\,4\,5\,6\,7\,8\,9}$ of the IIB diffeomorphism constraint is 
associated with the root space of $-\delta$, just as is the component 
${2\,3\,4\,5\,6\,7\,8\,9\,10}$ of the $D=11$ diffeomorphism constraint, see appendix~\ref{IIBapp}.
This suggests that, possibly the two expressions agree completely. Inspecting all the different root components 
and $E_{10}$ generators one verifies
\be\label{deltacon}
\cL{-3}{2\,3\,4\,5\,6\,7\,8\,9\,10}\Big|_{\text{ real roots}} = 
\Lb{-4}{2\,3\,4\,5\,6\,7\,8\,9}\Big|_{\text{ real roots}} = 
\frac13 \mc{L}_{-\delta}\Big|_{\text{ real roots}}\,,
\ee
{\it i.e.}, the expressions agree on the bilinear expressions involving 
two real root generators --- as was, in fact, guaranteed by our use 
of the Weyl group in the covariantization procedure. We find it remarkable 
that there is such an agreement between the constraints of two different 
physical theories expressed in the simple algebraic fashion~(\ref{sugcharges}).

However, considering the bilinear terms contributing to the two expressions,
 one finds that there are terms that differ, an explicit example 
can be found in appendix~\ref{Cartanapp}. One way to interpret this difference is
the following: The full set of constraints can be divided in two parts:
(i) a universal part, based on the `skeleton' of null roots, and comprising 
the `scaffold' of special configurations 
\be\label{skel}
\mf{L}_\alpha = \sum_{\beta_1+\beta_2=\alpha\atop \beta_1,\beta_2\,\text{real}} J_{\beta_1} J_{\beta_2}\quad\quad\text{for $\alpha$ null}
\ee
and, (ii) a non-universal part (the `flesh') 
that depends on the choice of subgroup
under which one covariantizes the `scaffold' part (\ref{skel}).

The universal part of the construction (\ref{skel}) has the property of being
 preserved by the action of the discrete Weyl group $W(E_{10})$ and its subgroups 
$W(A_9)$ and $W(A_8\oplus A_1)$. By contrast, the covariantization 
of 
the skeleton under the corresponding continuous groups $GL(10,\reals)$ (for $D=11$) and 
$GL(9,\reals) \times  SL(2,\reals)$ (for type IIB) leads 
to different results on the additional new terms inside the light-cone 
that are generated by the covariantization. That different new terms are 
possible is due to the fact that in those terms one has to specify the 
coefficients $M_{s_1,s_2}(\beta_1,\beta_2)$ for the contraction of root 
spaces of different dimensions.\footnote{A similar difference was 
  already noted for the $E_9$ contraction in~\cite{Damour:2007dt}.} These are fixed 
by covariance under a 
chosen level decomposition subgroup.\footnote{See, however, 
 the suggestion above, Eq. (\ref{sugadjoint}) 
 that one might use the Lie algebra generator
 associated  to the considered constraint-root $\alpha$ to define a universal way
 of pairing the two different root spaces $\mf{g_{\beta_1}}$, $\mf{g_{\beta_2}}$.}
 
We believe there is some evidence that hyperbolic algebras may admit
a realisation akin to the realisation of affine algebras in terms of a 
spectral parameter\footnote{Some evidence from the structure of the compact subgroup $K(E_{10})$ was given in~\cite{Kleinschmidt:2006dy}.}, but our results here strongly suggest that, if there 
is such a realisation, it will not be unique. Thinking of Sugawara 
constructions as being associated with spectral parameters, this can be 
interpreted by saying that the $E_{10}$ algebra may not possess a single 
or unique set of spectral parameters. Rather, one can and has to choose 
a set of spectral parameters by covariantizing under a subalgebra of 
one's choice. If the spectral parameters are related to space variables 
(as is the case for the affine algebras appearing in $D=2$ supergravities),
then this would be in good agreement 
with the anticipation that one can make space-times of different 
dimensions emergent from $E_{10}$, depending on the choice of level 
decomposition~\cite{Damour:2007bd}. From this point of view the hyperbolic Sugawara 
construction considered here is less unique than in the affine case 
since it depends on the choice of level decomposition. At the same time 
it nicely incorporates  the expected possibility of having different spaces emerging 
from an U-duality (Weyl group) invariant scaffold.

On the other hand, restricting only to real root generators, we can now 
use the agreement between the two expressions to construct new terms 
involving higher level generators, showing the full power of the approach.
The crucial point is that the IIB diffeomorphism constraint (\ref{IIBdiff}) also 
contains other components that are not contained in the previous expressions 
(\ref{oldconstraints}) corresponding to the $A_9$ level decomposition 
with the level truncation appropriate to $D=11$ supergravity, for example the real root combination
\be\label{IIBdiffparts}
\Lb{-4}{2\,3\,4\,5\,6\,7\,8\,9} \,\ni \,
\frac13 \Jb{-4}{8|2\,3\,4\,5\,6\,7\,8}\Jb{0}{9}{}_8\,.
\ee
Translating again between the two different bases of $E_{10}$ using
\be
 \Jb{0}{9}{}_8 = \J{+1}{}{}_{\!\!8\,9\,10}\,,\quad \Jb{-4}{8|2\,3\,4\,5\,6\,7\,8} = \J{-4}{8\,9\,10|2\,3\,4\,5\,6\,7\,8\,9\,10}\,.
\ee
we infer that this is part of an extended $\mf{sl}(10)$ covariant expression, namely
\be
\cL{-3}{m_1\ldots m_9} \rightarrow (\ref{olddiff}) \; + \;
\frac{1}{3\cdot 3!} \J{+1}{}_{\!\!\!p_1p_2p_3}\J{-4}{p_1p_2p_3|m_1\ldots m_9}\,.
\ee
The normalization is fixed by the term in the IIB expansion. This is also 
the only possible contraction between $A_9$ level $+1$ and $-4$ contributing 
to the diffeomorphism constraint in $D=11$. (The mass deformation generator 
on $\ell=4$ does not contribute to the diffeomorphism constraint~\cite{Henneaux:2008nr}.) We note 
that the generator appearing in this new piece of the $D=11$ 
diffeomorphism constraint is a gradient generator in the language 
of~\cite{Damour:2002cu}. The new term in the $D=11$ constraint now has 
components on IIB level $\ell=-5$ and $\ell=-6$ that can be covariantized now under 
$\mf{sl}(9)\oplus\mf{sl}(2)$ generating new terms. We have carried 
out this procedure one step farther and found the following expressions 
for the `diffeomorphism constraints' in $A_9$ decomposition
\be\label{A9diff}
\cL{-3}{m_1\ldots m_9} &=&28 \J{-1}{m_1m_2m_3}\J{-2}{m_4\ldots m_6} + 
3 \J{-3}{p|m_1\ldots m_8}\J{0}{m_9}{}_p\\ &&
\!\!\!\!\!\!\!\!\!\!\!\!\!\!\!\!\!\!\!\!
+\,\frac{1}{3\cdot 3!}\J{-4}{p_1p_2p_3|m_1\ldots m_9}\J{+1}{}{}_{\!\!\!p_1p_2p_3} + 
\frac{1}{3\cdot 6!} \J{-5}{p_1\ldots p_6|m_1\ldots m_9}\J{+2}{}{}_{\!\!\! p_1\ldots p_6} \nn\\
&& +  \, \dots \,.\nn
\ee
with implicit antisymmetrization over $[m_1\ldots m_9]$, and a 
corresponding expression in $A_8\oplus A_1$ decomposition
\be\label{IIBdiff2}
\Lb{-4}{m_1\ldots m_8}& =&\frac{35}{3}\Jb{-2}{m_1\ldots m_4}\Jb{-2}{m_5\ldots m_8}- 
\frac{28}{3}\Jb{-1}{m_1m_2,\alpha}\Jb{-3}{m_3\ldots m_8,\beta} \epsilon_{\alpha\beta} \nn\\
&& \!\!\!\!\!\!\!\!\!\!\!\!\!\!\!\!\!\!\!\!
-\, \frac13 \Jb{-4}{m_1\ldots m_8,\alpha\gamma}\Jb{0}{\beta}{}_\gamma\epsilon_{\alpha\beta}
- \frac{8}{3}\Jb{-4}{p|m_1\ldots m_7}\Jb{0}{m_8}{}_p \\
&&\!\!\!\!\!\!\!\!\!\!\!\!\!\!\!\!\!\!\!\!
+\, \frac1{3\cdot 2!} \Jb{-5}{p_1p_2|m_1\ldots m_8,\alpha}\Jb{+1}{}{}_{\!\!\!p_1p_2,\alpha} +\frac{1}{3\cdot 4!}\Jb{-6}{p_1\ldots p_4|m_1\ldots m_8}\Jb{+2}{}{}_{\!\!\! p_1\ldots p_4}\nn\\
&& + \dots \nn
\ee
with implicit antisymmetrization over $[m_1\ldots m_8]$. Note that the 
index range of the world indices is different in the two decompositions: In the $D=11$ case, corresponding to $A_9$, the index range  is $m=1,\ldots,10$ and in the type IIB case, corresponding to $A_8\oplus A_1$, the index range is $m=1,\ldots,9$.
By construction, these two expressions have the property that they agree 
on the real roots. In this way one produces an expression for a Sugawara 
constraint $\mc{L}_{-\delta}$ which extends to arbitrarily positive and negative 
step operators. 
We also note the appearance of {\em gradient generators}
precisely in accord with (\ref{sugaff}), as these generators are the
ones that reduce to the higher level affine generators in the truncation
of $E_{10}$ to $E_9$~\cite{Kleinschmidt:2006dy}. The gradient generators are those generators related to real roots of the affine $E_9$~\cite{Damour:2002cu,Nicolai:2003fw}.
It is straightforward to see that the infinite
prolongation of our procedure will give rise to all the terms needed
to match with the full sum in (\ref{sugaff}), for negative values of $n$. 

Our see-saw mechanism not only demands the extension of the 
constraints $\mf{L}_\alpha$ (for a given $\alpha$) to infinite strings
of bilinears of Noether charges in agreement with the affine Sugarawara 
construction, but {\em also allows to switch between constraints that are 
distinct as supergravity constraints}. For instance, certain components of the 
IIB diffeomorphism constraint metamorphose into components of the 
$D=11$ Gauss constraint when viewed in a different level decomposition! 
To see this more explicitly, consider the following component of 
the IIB diffeomorphism constraint (\ref{IIBdiff2})
\be
\Lb{-4}{1\,2\,3\,4\,5\,6\,7\,8} &\ni& \frac13 \Jb{-2}{1\,2\,3\,4}\Jb{-2}{5\,6\,7\,8} +\ldots
\ee
where we only picked out one real root combination for simplicity. Translating this to the $A_9$ basis via
\be
\Jb{-2}{1\,2\,3\,4} = \J{-2}{1\,2\,3\,4\,9\,10}\,,\quad \Jb{-2}{5\,6\,7\,8} = \J{-2}{5\,6\,7\,8\,9\,10}\,,
\ee
we find that it is part of a covariant expression
\be
\cL{-4}{m_1\ldots m_{10}||n_1n_2} = 42\J{-2}{n_1[m_1\ldots m_5}\J{-2}{m_6\ldots m_{10}]n_2} - (n_1\leftrightarrow n_2)
\ee
where the overall normalization differs by a factor of $20$ from (\ref{oldgauss}), see also (\ref{diffcomp}) in comparison to (\ref{gausscomp}). This is exactly the combination that appears in the Gauss constraint of $D=11$ supergravity, a result not too surprising from the point of view of U-duality. 
Evidently, this process could now be continued {\em ad libitum}.

One can similarly generate new terms for the Gauss constraint in $D=11$, given up to $\ell=3$ in (\ref{oldgauss}). Starting from the following components of the IIB diffeomorphism constraint
\be
\Lb{-4}{1\,2\,3\,4\,5\,6\,7\,8} \ni \frac13 \left(\Jb{-4}{1\,2\,3\,4\,5\,6\,7\,8,\dot{1}\dot{1}}\Jb{0}{\dot{2}}{}_{\dot{1}} 
\; +\;  \Jb{-4}{8|2\,3\,4\,5\,6\,7\,8}\Jb{0}{1}{}_8
\right)\, +\, \ldots\,.
\ee
They can be mapped to $A_9$ quantities using the two distinct $A_9$ level $\ell=4$ representations
\be
\Jb{-4}{1\,2\,3\,4\,5\,6\,7\,8,\dot{1}\dot{1}} = \J{-4}{9|9|1\,2\,3\,4\,5\,6\,7\,8\,9\,10}\,, \quad
\Jb{-4}{8|2\,3\,4\,5\,6\,7\,8} = \J{-4}{8\,9\,10|2\,3\,4\,5\,6\,7\,8\,9\,10}\nn\\
\ee
to give $\mf{sl}(10)$ covariant additions to (\ref{oldgauss}) via
\be
\cL{-4}{m_1\ldots m_{10}||n_1n_2} &\rightarrow& (\ref{oldgauss}) 
\; + \;  \frac23\J{-4}{m_1\ldots m_{10}|p[n_1}\J{0}{n_2]}{}_p\nn\\
&&\quad 
+ \; \frac{10}{3}\,\J{-4}{m_1\ldots m_9|n_1n_2p}\J{0}{m_{10}}{}_p\, +\ldots\,.
\ee
We note that here both the gradient {\em and} non-gradient generator 
on $A_9$ level $\ell=4$ contribute. Since the mass deformation parameter 
of massive type IIA is contained in the non-gradient generator, this is 
in agreement with the fact that the Gauss constraint of massive IIA 
gets modified by the Romans mass~\cite{Romans:1985tz,Henneaux:2008nr}.

\end{subsection}

\begin{subsection}{General remarks on the construction}
\label{Remsec}

Let us summarize the construction and comment on some open questions
concerning this procedure. Starting from a single constraint $\mc{L}_{-\delta}$, associated to the primitive null root, we construct the `scaffold' as in (\ref{skel}), based on the decomposition $\alpha=\beta_1+\beta_2$ for $\alpha$ null and $\beta_1,\beta_2$ real.\footnote{We note that decompositions of imaginary roots into real roots have been considered in a different context in~\cite{Brown:2004jb}.} 
Since all root spaces involved are real, they 
are one-dimensional and there is no ambiguity in the contraction. There are also no ordering ambiguities at this level. We can then act on the expression (\ref{skel}) with $W(E_{10})$ to generate similar expressions for all null roots. This constitutes the full scaffold of the hyperbolic Sugawara constraints 
which is invariant (only) under $W(E_{10})$. The constraints of the type in (\ref{skel}) are both infinite in number and each consists of an infinite number of bilinears in the current components.

In order to construct constraints for the full $E_{10}$ one then needs to choose a level decomposition under a regular, finite-dimensional subalgebra. Covariance under this subalgebra induces additional terms on top of those already contained in the skeleton. The precise form of these additional terms depends on the subalgebra one chose in a systematic way, as is apparent from the explicit expressions in appendix~\ref{Cartanapp}. 
{}From that point of view it is clear that our construction is not covariant with respect to the full
 $E_{10}$ Lie algebra, but involves only the Weyl group $\mc{W}(E_{10})$ in a canonical way. Everything beyond that depends on the chosen subalgebra for the level decomposition.\footnote{In some sense this is also true for the affine $E_9$ Sugawara construction which uses as choice of subalgebra for the level decomposition $E_8$.} 
 
 We can also bring out the lack of `$E_{10}$ covariance'  by relating our construction to the question of an $E_{10}$ representation structure in the bilinear expression in $E_{10}$ generators. 
As already pointed out in footnote~\ref{hstfn} one might have liked to
identify the constraint $\mc{L}_{-\delta}$ with a
 highest weight vector of an integrable $E_{10}$ representation
with highest weight $\Lambda_1=-\delta$. If this were the case the constraint 
should be annihilated by all raising operators. Here, we recall that we 
express the step operators in terms of current components (rather than
in terms of the `contragredient' $E_{10}$ Lie algebra generators 
$T_\alpha^{(s)}$), so that for example $e_1=J^2{}_1$. Using the explicit 
expression for $\mc{L}_{-\delta}$ in $A_9$ decomposition we find that
\be
\left[ e_i , \cL{-3}{2\,3\,4\,5\,6\,7\,8\,9\,10} \right] = 
\lb \J{0}{i+1}{}_{i}\, ,\cL{-3}{2\,3\,4\,5\,6\,7\,8\,9\,10}\rb =0 
\quad\text{for $i=1,\ldots,9$}
\ee
where all commutators should be read as Poisson (or Dirac) brackets
in the canonical setting. However, for the $e$ generator corresponding 
to the omitted node we get
\be\label{A9fail}
\left[ e_{10} , \cL{-3}{2\,3\,4\,5\,6\,7\,8\,9\,10} \right] =\lb \J{1}{}{}_{\!8\,9\,10},\cL{-3}{2\,3\,4\,5\,6\,7\,8\,9\,10}\rb \neq 0 \,,
\ee
showing that this component of the constraint generator {\em is only
a highest weight state with respect to the $A_9$ subalgebra, but not the 
full $E_{10}$ algebra}. Since the $A_9$ expression agrees with the $A_8\oplus A_1$ 
expression on the real roots, we can repeat the calculation in IIB variables to find
\be
\left[ e_i , \Lb{-4}{2\,3\,4\,5\,6\,7\,8\,9} \right] = 0 \quad\text{for $i=1,\ldots,7,9,10$}
\ee
and
\be\label{IIBfail}
\left[ e_{8} , \Lb{-4}{2\,3\,4\,5\,6\,7\,8\,9} \right] \neq 0 \,,
\ee
Being a highest weight vector now with respect to $A_8\oplus A_1$, this
is different from the result for the $A_9$ decomposition, but again
illustrates the lack of full $E_{10}$ covariance. 

Similar conclusions hold for the $D_9\equiv SO(9,9)$ decomposition 
of~\cite{Kleinschmidt:2004dy}. Without going into the details of the 
calculation, the lowest order constraint for the $D_9$ decomposition is
\be\label{D9}
\cL{-2}{I} = \frac12\J{0}{KL} \J{-2}{IKL} +  \frac12\J{-1}{A} ({\mc{C}}\Gamma^I)_{AB} \J{-1}{B} + \dots
\ee
where $I,K,L=1,\ldots,18$ and $A,B=1,\ldots,256$ are $SO(9,9)$ vector and spinor indices, respectively,
and we use again the symbol $J$ to denote the components of the conserved 
$E_{10}$ current, but now in the $D_9$ decomposition. The 18=9+9 constraints
in (\ref{D9}) correspond to the diffeomorphism constraint and the Gauss constraint
for the Neveu-Schwarz 2-form field of IIA theory; alternatively, they might be
interpreted as a doubled set of diffeomorphism constraints w.r.t. the
nine spatial target space coordinates $X^i$ and their  (world-sheet)
`duals' $\tilde{X}^i$~\cite{Kleinschmidt:2004dy}. As before it is the omitted node ({\it i.e.}, node 9 for 
the massive IIA theory) which causes failure of the construction: by $SO(9,9)$ covariance,
the dilaton field associated with this node cannot appear in (\ref{D9}). 
Accordingly, it is now the generator $e_9$ which does not annihilate 
the relevant component of $\mf{L}^{(-2)}$. 

To summarize: The failure of the constraint to be a highest weight 
vector w.r.t. the full $E_{10}$ algebra is invariably associated with 
the node that has been deleted for the given level decomposition. In
Appendix~\ref{Cartanapp} we show that a related statement applies to the dependence
of the constraints on the Cartan subalgebra generators.

One further interesting aspect of our construction is that, to start with, it associates a constraint with every Weyl image of the fundamental null root $- \delta$. In the same way one can associate constraints to the Weyl images of $-n\delta$ and in this way obtains a constraint for every $E_{10}$ root on the (past) light-cone. After choosing a level decomposition subalgebra one generates additional constraints {\em inside} the light-cone by covariantization under this subalgebra. 

It is possible that, as indicated in subsection 3.1.3, the set of roots $\mc{C}$ `supporting'
the full set of constraints be {\em universally} given by all the weights inside the (past) light-cone. This set can also be described as the union of the weight diagrams of the 
representations $L(\Lambda_1)$, $L(\Lambda_1)\otimes L(\Lambda_1)$,   $L(\Lambda_1)\otimes L(\Lambda_1)\otimes L(\Lambda_1)$, etc..
On the other hand, the precise 
Sugawara-like expression defining the constraint $\mf{L}_{\alpha}$ associated to some
$\alpha \in \mc{C}$ seems to depend on the choice of a level decomposition.

Finally, note that since we are defining an infinity of constraints associated with all null roots of the hyperbolic algebra $E_{10}$, one might worry whether there are any solutions that satisfy the geodesic equation and all the Sugawara constraints. It is reassuring to note that there are such solutions, namely for example the Kasner cosmologies. These correspond to only non-vanishing Cartan subalgebra components of the current and hence all constraints except the Hamiltonian constraint (\ref{hcsug}) are trivially satisfied. Other solutions correspond to specific cases of Bianchi cosmologies. The exact count of the remaining number of degrees of freedom is quite involved and beyond the scope of this paper.

\end{subsection}

\end{section}
\vspace{5mm}

\noindent{\bf Acknowledgements.}
We would like to thank Ofer Gabber and Victor Kac for informative discussions.
AK is a Research Associate of the Fonds de la Re\-cherche--FNRS, Belgium, and would like to thank IHES and AEI for hospitality. This work has 
been supported in part by IISN-Belgium (conventions 
4.4511.06, 4.4505.86 and 4.4514.08) and by the Belgian Federal Science 
Policy Office through the Interuniversity Attraction Pole P6/11. 

\appendix

\begin{section}{Level decompositions}
\label{ldapp}

For the reader's convenience, we collect in this appendix some results
on the level decompositions of $E_{10}$ appropriate for $D=11$ supergravity 
and for type IIB in $D=10$. These appeared originally 
in~\cite{Damour:2002cu,Nicolai:2003fw} and~\cite{Kleinschmidt:2003mf,Kleinschmidt:2004rg}, respectively.

\begin{subsection}{Level decomposition under $A_9$}
\label{a9ld}

The $A_9\cong \mf{sl}(10)$ subalgebra relevant for $D=11$ supergravity is obtained by removing node~10 from the Dynkin diagram of fig.~\ref{e10dynk}.

\begin{longtable}{|c|c|c|c|c|}
\hline
$\ell$ & $A_9$ Dynkin labels & $E_{10}$ root for lowest weight & $\mu$ & $\alpha^2$  \\
\hline\hline\endhead
0&[1,0,0,0,0,0,0,0,1]&(-1,-1,-1,-1,-1,-1,-1,-1,-1,0) & 1&2\\
0&[0,0,0,0,0,0,0,0,0]&(0,0,0,0,0,0,0,0,0,0)& 1&0\\
1&[0,0,0,0,0,0,1,0,0]&(0,0,0,0,0,0,0,0,0,1)&	1&	2\\
2&[0,0,0,1,0,0,0,0,0]&(0,0,0,0,1,2,3,2,1,2)&	1&	2\\
3&[0,1,0,0,0,0,0,0,1]&(0,0,1,2,3,4,5,3,1,3)&	1&	2\\
3&[1,0,0,0,0,0,0,0,0]&(0,1,2,3,4,5,6,4,2,3)&	0&	0\\
4&[0,0,0,0,0,0,0,0,2]&(1,2,3,4,5,6,7,4,1,4)&	1&	2\\
4&[0,0,0,0,0,0,0,1,0]&(1,2,3,4,5,6,7,4,2,4)&	0&	0\\
4&[1,0,0,0,0,0,1,0,0]&(0,1,2,3,4,5,6,4,2,4)&	1&	2\\
5&[0,0,0,0,0,1,0,0,1]&(1,2,3,4,5,6,8,5,2,5)&	1&	2\\
5&[0,0,0,0,1,0,0,0,0]&(1,2,3,4,5,7,9,6,3,5)&	0&	0\\
5&[1,0,0,1,0,0,0,0,0]&(0,1,2,3,5,7,9,6,3,5)&	1&	2\\
6&[0,0,0,1,0,0,0,1,0]&(1,2,3,4,6,8,10,6,3,6)&	1&	2\\
6&[0,0,1,0,0,0,0,0,1]&(1,2,3,5,7,9,11,7,3,6)&	1&	0\\
6&[0,1,0,0,0,0,0,0,0]&(1,2,4,6,8,10,12,8,4,6)&	1&	-2\\
6&[1,1,0,0,0,0,0,0,1]&(0,1,3,5,7,9,11,7,3,6)&	1&	2\\
6&[2,0,0,0,0,0,0,0,0]&(0,2,4,6,8,10,12,8,4,6)&	0&	0\\
7&[0,0,0,0,0,0,0,0,1]&(2,4,6,8,10,12,14,9,4,7)&	1&	-4\\
7&[0,0,1,0,0,1,0,0,0]&(1,2,3,5,7,9,12,8,4,7)&	1&	2\\
7&[0,1,0,0,0,0,0,1,1]&(1,2,4,6,8,10,12,7,3,7)&	1&	2\\
7&[0,1,0,0,0,0,1,0,0]&(1,2,4,6,8,10,12,8,4,7)&	1&	0\\
7&[1,0,0,0,0,0,0,0,2]&(1,3,5,7,9,11,13,8,3,7)&	1&	0\\
7&[1,0,0,0,0,0,0,1,0]&(1,3,5,7,9,11,13,8,4,7)&	2&	-2\\
7&[2,0,0,0,0,0,1,0,0]&(0,2,4,6,8,10,12,8,4,7)&	1&	2\\
8&[0,0,0,0,0,0,0,1,2]&(2,4,6,8,10,12,14,8,3,8)&	1&	2\\
8&[0,0,0,0,0,0,0,2,0]&(2,4,6,8,10,12,14,8,4,8)&	0&	0\\
8&[0,0,0,0,0,0,1,0,1]&(2,4,6,8,10,12,14,9,4,8)&	2&	-2\\
8&[0,0,0,0,0,1,0,0,0]&(2,4,6,8,10,12,15,10,5,8)&	2&	-4\\
8&[0,0,2,0,0,0,0,0,0]&(1,2,3,6,9,12,15,10,5,8)&	1&	2\\
8&[0,1,0,0,1,0,0,0,1]&(1,2,4,6,8,11,14,9,4,8)&	1&	2\\
8&[0,1,0,1,0,0,0,0,0]&(1,2,4,6,9,12,15,10,5,8)&	1&	0\\
8&[1,0,0,0,0,0,1,1,0]&(1,3,5,7,9,11,13,8,4,8)&	1&	2\\
8&[1,0,0,0,0,1,0,0,1]&(1,3,5,7,9,11,14,9,4,8)&	2&	0\\
8&[1,0,0,0,1,0,0,0,0]&(1,3,5,7,9,12,15,10,5,8)&	2&	-2\\
8&[2,0,0,1,0,0,0,0,0]&(0,2,4,6,9,12,15,10,5,8)&	1&	2\\\hline
\end{longtable}

The low-lying generators are denoted by ($a,b,\ldots=1,\ldots,10$)
\be
\ell=0 &:& K^{a}{}_{b}\nn\\
\ell=1 &:& E^{abc} = E^{[abc]}\nn\\
\ell=2 &:& E^{a_1\ldots a_6} = E^{[a_1\ldots a_6]}\nn\\
\ell=3 &:& E^{a_0|a_1\ldots a_8} = E^{a_0|[a_1\ldots a_8]}\\
\ell=4 &:& E^{a_1a_2a_3|b_1\ldots b_9} = E^{[a_1a_2a_3]|[b_1\ldots b_9]}\quad\text{and}\quad
 E^{a|b|c_1\ldots c_{10}} = E^{(a|b)|[c_1\ldots c_{10}]}\nn
 \ee
(with the usual irreducibility conditions $E^{[a_0|a_1\ldots a_8]} = 0$,
etc.). They are related to the Chevalley--Serre generators by
\be\label{a9cs1} 
e_i = K^i{}_{i+1}\,,\quad f_i = K^{i+1}{}_i\,,\quad h_i = K^i{}_i -K^{i+1}{}_{i+1}\quad(i=1,\ldots,9)
\ee
and
\be\label{a9cs2}
e_{10} &=& E^{8\,9\,10}\,,\quad f_{10} = F_{8\,9\,10}\,,\nn\\
h_{10} &=& -\frac13K+ K^{8}{}_8+K^9{}_9 + K^{10}{}_{10}\,,
\ee
where $K=\sum_{a=1}^{10} K^a{}_a$. Commutation relations for these generators can be found in~\cite{Damour:2004zy,Henneaux:2008nr} but note that we have rescaled all generators such that their lowest weight elements (e.g. $E^{10|3\,4\,5\,6\,7\,8\,9\,10}$) have norm $1$.

\end{subsection}

\begin{subsection}{Level decomposition under $A_8\oplus A_1$}
\label{a8a1ld}

The $A_8\oplus A_1\cong \mf{sl}(9)\oplus\mf{sl}(2)$ subalgebra relevant for type IIB supergravity is obtained by removing node~8 from the Dynkin diagram of fig.~\ref{e10dynk}.

\begin{longtable}{|c|c|c|c|c|}
\hline
$\ell$ & $A_8\oplus A_1$ Dynkin labels & $E_{10}$ root for lowest weight & $\mu$ & $\alpha^2$  \\
\hline\hline\endhead
0&	[1,0,0,0,0,0,0,1][0]&(-1,-1,-1,-1,-1,-1,-1,0,0,-1)&1&2\\
0&	[0,0,0,0,0,0,0,0][0]&(0,0,0,0,0,0,0,0,0,0)&1&0\\
0&	[0,0,0,0,0,0,0,0][2]&(0,0,0,0,0,0,0,0,-1,0)&1&2\\
1&	[0,0,0,0,0,0,1,0][1]&(0,0,0,0,0,0,0,1,0,0)&	1&2	\\
2&	[0,0,0,0,1,0,0,0][0]&(0,0,0,0,0,1,2,2,1,1)&	1&2\\
3&	[0,0,1,0,0,0,0,0][1]&(0,0,0,1,2,3,4,3,1,2)&	1&2\\
4&	[0,1,0,0,0,0,0,1][0]&(0,0,1,2,3,4,5,4,2,2)&	1&2\\
4&	[1,0,0,0,0,0,0,0][0]&(0,1,2,3,4,5,6,4,2,3)&	0&0\\
4&	[1,0,0,0,0,0,0,0][2]&(0,1,2,3,4,5,6,4,1,3)&	1&2\\
5&	[0,0,0,0,0,0,0,1][1]&(1,2,3,4,5,6,7,5,2,3)&	1&0\\
5&	[1,0,0,0,0,0,1,0][1]&(0,1,2,3,4,5,6,5,2,3)&	1&2\\
6&	[0,0,0,0,0,0,1,1][0]&(1,2,3,4,5,6,7,6,3,3)&	1&2\\
6&	[0,0,0,0,0,1,0,0][0]&(1,2,3,4,5,6,8,6,3,4)&	0&0\\
6&	[0,0,0,0,0,1,0,0][2]&(1,2,3,4,5,6,8,6,2,4)&	1&2\\
6&	[1,0,0,0,1,0,0,0][0]&(0,1,2,3,4,6,8,6,3,4)&	1&2\\
7&	[0,0,0,0,1,0,0,1][1]&(1,2,3,4,5,7,9,7,3,4)&	1&2\\
7&	[0,0,0,1,0,0,0,0][1]&(1,2,3,4,6,8,10,7,3,5)&	1&0\\
7&	[1,0,1,0,0,0,0,0][1]&(0,1,2,4,6,8,10,7,3,5)&	1&2\\
8&	[0,0,0,1,0,0,1,0][0]&(1,2,3,4,6,8,10,8,4,5)&	1&2\\
8&	[0,0,1,0,0,0,0,1][0]&(1,2,3,5,7,9,11,8,4,5)&	1&0\\
8&	[0,0,1,0,0,0,0,1][2]&(1,2,3,5,7,9,11,8,3,5)&	1&2\\
8&	[0,1,0,0,0,0,0,0][0]&(1,2,4,6,8,10,12,8,4,6)& 2&-2\\
8&	[0,1,0,0,0,0,0,0][2]&(1,2,4,6,8,10,12,8,3,6)& 1&0\\
8&	[1,1,0,0,0,0,0,1][0]&(0,1,3,5,7,9,11,8,4,5)&	1&2\\
8&	[2,0,0,0,0,0,0,0][0]&(0,2,4,6,8,10,12,8,4,6)& 0&0\\
8&	[2,0,0,0,0,0,0,0][2]&(0,2,4,6,8,10,12,8,3,6)& 1&2\\\hline
\end{longtable}
\allowdisplaybreaks{
The low-lying generators are (now $a,b,\ldots=1,\ldots,9$ are $\mf{sl}(9)$ vector indices and $\alpha,\beta=\dot{1},\dot{2}$ are $\mf{sl}(2)$ vector indices)
\be
\ell=0 &:& K^{a}{}_b\quad \text{and}\quad K^{\alpha}{}_{\beta}\quad\text{(with $\delta^\beta_\alpha K^\alpha{}_\beta = 0$)}\nn\\
\ell=1 &:& E^{ab,\alpha}=E^{[ab],\alpha}\nn\\
\ell=2 &:& E^{a_1a_2a_3a_4} = E^{[a_1a_2a_3a_4]}\nn\\
\ell=3 &:& E^{a_1\ldots a_6,\alpha} = E^{[a_1\ldots a_6],\alpha}\\
\ell=4 &:& E^{a_0|a_1\ldots a_7} = E^{a_0|[a_1\ldots a_7]}\quad\text{and}\quad E^{a_1\ldots a_8,\alpha\beta}=E^{[a_1\ldots a_8],(\alpha\beta)}\nn
\ee
The relation to the Chevalley--Serre generators is now given by}
\begin{align}\label{IIBCS1}
e_i &= K^i{}_{i+1}\,,&\quad f_i& = K^{i+1}{}_i\,,&\quad h_i &= K^i{}_i -K^{i+1}{}_{i+1}\quad(i=1,\ldots,7)&\nn\\
e_{10} &= K^{8}{}_9\,,&\quad f_{10} &= K^9{}_8\,,&\quad h_{10}&=K^8{}_8-K^9{}_9\,,&\nn\\
e_9 &=K^{\dot{1}}{}_{\dot{2}}\,,&\quad f_9 &=K^{\dot{2}}{}_{\dot{1}}\,,&\quad h_9&=K^{\dot{1}}{}_{\dot{1}}-K^{\dot{2}}{}_{\dot{2}}\,.&
\end{align}
The explicit dots on the indices indicate numerical values for $\mf{sl}(2)$ vector indices. For the deleted node $8$ one has
\be\label{IIBCS2}
e_{8} &=& E^{8\,9,\dot{2}}\,,\quad f_{8} = F_{8\,9,\dot{2}}\,,\nn\\
h_{8} &=& -\frac14K+ K^{8}{}_8 + K^{9}{}_{9}-\frac12\left(K^{\dot{1}}{}_{\dot{1}}-K^{\dot{2}}{}_{\dot{2}}\right)\,,
\ee
where now $K=\sum_{a=1}^9 K^a{}_a$ is the trace in $\mf{gl}(9)$. Commutation relations for these generators can be found in~\cite{Kleinschmidt:2004rg}, where we used an $\mf{so}(1,2)$ spinor and vector notation instead of $\mf{sl}(2)$ tensors as above.

\end{subsection}

\end{section}

\begin{section}{Constraints of type IIB supergravity and universality}
\label{IIBapp}

The Einstein equation of motion of IIB supergravity can be written as
\be
R_{AB} &=& -\frac14 S_A^\alpha S_{B,\alpha} + \frac1{96} F_{A}{}^{C_1\ldots C_4} F_{BC_1\ldots C_4}\nn\\
&&+\frac14 H_{A}{}^{C_1C_2,\alpha} H_{BC_1C_2,\alpha} -\frac1{48}\eta_{AB}H^{C_1\ldots C_3,\alpha}H_{C_1C_2C_3\alpha}
\ee
in flat indices, where we corrected a factor of two compared 
to~\cite{Kleinschmidt:2004rg}.  

The diffeomorphism constraint is obtained as the $0a$ component of this equation. Using self-duality of $F$ and the dictionary of~\cite{Kleinschmidt:2004rg} one finds, up to overall normalization, the expression
\be
\Lb{-4}{m_1\ldots m_8}& =&\frac{35}{3}\Jb{-2}{m_1\ldots m_4}\Jb{-2}{m_5\ldots m_8}+ \frac{28}{3}\Jb{-1}{m_1m_2,\alpha}\Jb{-3}{m_3\ldots m_8,\beta} \epsilon_{\alpha\beta} \nn\\
&& +\frac13 \Jb{-4}{m_1\ldots m_8,\alpha\gamma}\Jb{0}{\beta}{}_\gamma\epsilon_{\alpha\beta}
+ \frac{8}{3}\Jb{-4}{p|m_1\ldots m_7}\Jb{0}{m_8}{}_p 
\ee
in terms of the $E_{10}$ current components in $A_8\oplus A_1$ decomposition.

\end{section}

\begin{section}{Explicit expressions involving Cartan generators}
\label{Cartanapp}

In this appendix, we give explicit expressions for the contractions between 
the Cartan subalgebra and the $\delta$ root space to show that the $A_9$ 
and $A_8\oplus A_1$ covariant expressions (\ref{A9diff}) and (\ref{IIBdiff2}) 
differ, thereby illustrating that (\ref{deltacon}) is indeed only valid 
on contractions of real root spaces.

Consider the contributions from the Cartan subalgebra to the highest component of (\ref{A9diff}). They come exclusively from the $J^{(-3)} J^{(0)}$ contraction and are
\be
3\cL{-3}{2\,3\,4\,5\,6\,7\,8\,9\,10} &\ni&\J{-3}{2|3\,4\,5\,6\,7\,8\,9\,10}\J{0}{2}{}_{2}+\J{-3}{3|4\,5\,6\,7\,8\,9\,10\,2}\J{0}{3}{}_{3}\nn\\
&&+\dots+\J{-3}{10|2\,3\,4\,5\,6\,7\,8\,9}\J{0}{10}{}_{10}\\
&&
\!\!\!\!\!\!\!\!\!\!\!\!\!\!\!\!\!\!\!\!\!\!\!\!\!\!\!\!\!\!\!\!\!\!\!\!\!\!\!\!\!\!\!\!\!\!\!
= \J{-3}{2|3\,4\,5\,6\,7\,8\,9\,10} (h_2+h_3+\ldots+h_8+ h_9) + \ldots+ \J{-3}{9|10\,2\,3\,4\,5\,6\,7\,8}h_9\nn
\ee
where the hook symmetry
\be
\J{-3}{[2|3\,4\,5\,6\,7\,8\,9\,10]} = 0
\ee
of the level three element was used and we identified for simplicity the current component with the corresponding Cartan generators using (\ref{a9cs1}) and (\ref{a9cs2}). We see that only the Cartan generators  of the $A_9$ `gravity line' appear in this contraction. The only `missing' ones are the one from the deleted node $10$ and the hyperbolic node $1$. The latter is related to our choice 
of (highest) component.

Repeating the same calculation for the $A_8\oplus A_1$ decomposition and (\ref{IIBdiff2}) one finds similarly
\be
3\Lb{-4}{2\,3\,4\,5\,6\,7\,8\,9} &\ni& \ 
\Jb{-4}{2|3\,4\,5\,6\,7\,8\,9}\Jb{0}{2}{}_2 \, + \ldots + \, 
\Jb{-4}{9|2\,3\,4\,5\,6\,7\,8}\Jb{0}{9}{}_9\nn\\
&&+ \Jb{-4}{2\,3\,4\,5\,6\,7\,8\,9,\dot{1}\dot{2}}
(\Jb{0}{\dot{1}}{}_{\dot{1}}-\Jb{0}{\dot{2}}{}_{\dot{2}})\\
&=& \Jb{-4}{2|3\,4\,5\,6\,7\,8\,9}(h_2+\ldots+h_7+h_{10}) + \Jb{-4}{8|9\,2\,3\,4\,5\,6\,7} h_{10}\nn\\
&&+\Jb{-4}{2\,3\,4\,5\,6\,7\,8\,9,\dot{1}\dot{2}} h_9\nn
\ee
The Cartan generators that appear in this expression are those from 
the $A_8\oplus A_1$ gravity line, so the `missing' generators are that 
of the deleted node $8$ and of the hyperbolic node $1$. The latter is 
again related to our choice of component of the diffeomorphism 
constraint so that the real discrepancy between the two expressions 
can be traced again to the different deleted nodes. This is related 
to the failure of this constraint to be a highest weight vector, 
see the expressions (\ref{A9fail}) and (\ref{IIBfail}).

Finally, it is clear that the Cartan generator missing in (\ref{D9}) is 
$h_9$, as the diagonal generators among the $SO(9,9)$ generators 
$J^{(0)\,KL}$ are identified with $h_1,\dots,h_8, h_{10}$, while 
$h_9$ is associated with the dilaton, again confirming our general conclusion.

\end{section}


\begin{thebibliography}{40}

\bibitem{DeWitt:1967yk}
  B.~S.~DeWitt,
  {\sl Quantum Theory of Gravity. 1. The Canonical Theory},
  Phys.\ Rev.\  {\bf 160} (1967) 1113.

\bibitem{Kiefer:2004gr}
  C.~Kiefer,
  {\sl Quantum gravity},
  Int.\ Ser.\ Monogr.\ Phys.\  {\bf 124} (Oxford University Press, 2004).
  
\bibitem{Damour:2002cu}
  T.~Damour, M.~Henneaux and H.~Nicolai,
  {\sl $E_{10}$ and a 'small tension expansion' of M Theory},
  Phys.\ Rev.\ Lett.\  {\bf 89} (2002) 221601
  [arXiv:hep-th/0207267].

\bibitem{Damour:2007dt}
  T.~Damour, A.~Kleinschmidt and H.~Nicolai,
  {\sl Constraints and the $E_{10}$ Coset Model},
  Class.\ Quant.\ Grav.\  {\bf 24} (2007) 6097
  [arXiv:0709.2691 [hep-th]].

\bibitem{Sugawara:1967rw}
  H.~Sugawara,
  {\sl A Field theory of currents},
  Phys.\ Rev.\  {\bf 170} (1968) 1659.

\bibitem{Bardakci:1970nb}
  K.~Bardak\c{c}i and M.~B.~Halpern,
  {\sl New dual quark models},
  Phys.\ Rev.\  D {\bf 3} (1971) 2493.

\bibitem{Goddard:1986bp}
  P.~Goddard and D.~I.~Olive,
  {\sl Kac-Moody And Virasoro Algebras In Relation To Quantum Physics},
  Int.\ J.\ Mod.\ Phys.\  A {\bf 1} (1986) 303.

\bibitem{KKN}
  A.~Kleinschmidt, M.~Koehn and H.~Nicolai,
  {\sl Supersymmetric quantum cosmological billiards},
  Phys. \ Rev. D {\bf 80} (2009) 061701
  [arXiv:0907.3048[gr-qc]]

\bibitem{Forte:2008jr}
  L.~A.~Forte,
  {\sl Arithmetical Chaos and Quantum Cosmology},
  Class.\ Quant.\ Grav.\  {\bf 26} (2009) 045001
  [arXiv:0812.4382 [gr-qc]].

\bibitem{Goddard:1972iy}
  P.~Goddard and C.~B.~Thorn,
 {\sl Compatibility of the Dual Pomeron with Unitarity and 
  the Absence of Ghosts in the Dual Resonance Model},
  Phys.\ Lett.\  B {\bf 40} (1972) 235.

\bibitem{Kleinschmidt:2004dy}
  A.~Kleinschmidt and H.~Nicolai,
  {\sl $E_{10}$ and $SO(9,9)$ invariant supergravity}
  JHEP {\bf 0407} (2004) 041
  [arXiv:hep-th/0407101].

\bibitem{Damour:2004zy}
  T.~Damour and H.~Nicolai,
  {\sl Eleven dimensional supergravity and the $E_{10}/K(E_{10})$ sigma-model at  low
  $A_9$ levels},
  arXiv:hep-th/0410245.

\bibitem{Kleinschmidt:2004rg}
  A.~Kleinschmidt and H.~Nicolai,
  {\sl IIB supergravity and $E_{10}$},
  Phys.\ Lett.\  B {\bf 606} (2005) 391
  [arXiv:hep-th/0411225].

 \bibitem{Henneaux:2008nr}
  M.~Henneaux, E.~Jamsin, A.~Kleinschmidt and D.~Persson,
  {\sl On the $E_{10}$/Massive Type IIA Supergravity Correspondence},
  Phys.\ Rev.\  D {\bf 79} (2009) 045008
  [arXiv:0811.4358 [hep-th]].
  
\bibitem{NS} H.~Nicolai and H.A.J.~Samtleben,
  {\sl On $K(E_9)$}, Q.J. Pure Appl. Math {\bf 1} (2005) 180,
  [hep-th/0407055]

\bibitem{West:2001as}
  P.~C.~West,
  {\sl $E_{11}$ and M theory},
  Class.\ Quant.\ Grav.\  {\bf 18} (2001) 4443
  [arXiv:hep-th/0104081].

\bibitem{West2003} P.~C.~West,
 {\sl $E(11), SL(32)$ and central charges},
 Phys.\ Lett.\ {\bf B575} (2003) 333
 [arXiv: hep-th/0307098]

\bibitem{Riccioni:2009hi}
  F.~Riccioni and P.~West,
  {\sl Local $E_{11}$},
  JHEP {\bf 0904} (2009) 051
  [arXiv:0902.4678 [hep-th]].

\bibitem{Schnakenburg:2001ya}
  I.~Schnakenburg and P.~C.~West,
  {\sl Kac-Moody symmetries of IIB supergravity},
  Phys.\ Lett.\  B {\bf 517} (2001) 421
  [arXiv:hep-th/0107181].
  
\bibitem{Schnakenburg:2002xx}
  I.~Schnakenburg and P.~C.~West,
  {\sl Massive IIA supergravity as a non-linear realisation},
  Phys.\ Lett.\  B {\bf 540} (2002) 137
  [arXiv:hep-th/0204207].

\bibitem{Kleinschmidt:2003mf}
  A.~Kleinschmidt, I.~Schnakenburg and P.~C.~West,
  {\sl Very-extended Kac-Moody algebras and their interpretation at low  levels},
  Class.\ Quant.\ Grav.\  {\bf 21} (2004) 2493
  [arXiv:hep-th/0309198].

\bibitem{West2004}
  P.~C.~West,
  {\sl The IIA, IIB and eleven-dimensional theories and their 
  common $E(11)$ origin},
  Nucl.\ Phys.\ {\bf B693} (2004) 76
  [arXiv: hep-th/0402140]

\bibitem{Kac} V.~G.~Kac, {\sl Infinite dimensional Lie algebras} (Cambridge University Press, 1995).

\bibitem{Damour:2002fz}
  T.~Damour, S.~de Buyl, M.~Henneaux and C.~Schomblond,
  {\sl Einstein billiards and overextensions of finite-dimensional simple Lie algebras},
  JHEP {\bf 0208} (2002) 030
  [arXiv:hep-th/0206125].

\bibitem{Gebert:1996vv}
  R.~W.~Gebert and H.~Nicolai,
  {\sl An affine string vertex operator construction at arbitrary level},
  J.\ Math.\ Phys.\  {\bf 38} (1997) 4435
  [arXiv:hep-th/9608014].

\bibitem{Gaberdiel:2002db}
  M.~R.~Gaberdiel, D.~I.~Olive and P.~C.~West,
  {\sl A class of Lorentzian Kac-Moody algebras},
  Nucl.\ Phys.\  B {\bf 645} (2002) 403
  [arXiv:hep-th/0205068].
    
\bibitem{Nicolai:2003fw}
  H.~Nicolai and T.~Fischbacher,
  {\sl Low level representations for $E_{10}$ and $E_{11}$},
  Cont. Math. {\bf 343} 191 (American Mathematical Society, 2004)
  [arXiv:hep-th/0301017].
  
\bibitem{Damour:2002et}
  T.~Damour, M.~Henneaux and H.~Nicolai,
  {\sl Cosmological billiards}
  Class.\ Quant.\ Grav.\  {\bf 20}, R145 (2003)
  [arXiv:hep-th/0212256].

\bibitem{Damour:2001sa}
  T.~Damour, M.~Henneaux, B.~Julia and H.~Nicolai,
  {\sl Hyperbolic Kac-Moody algebras and chaos in Kaluza-Klein models},
  Phys.\ Lett.\  B {\bf 509}, 323 (2001)
  [arXiv:hep-th/0103094].

\bibitem{KMW} V.~Kac, R.V.~Moody and M.~Wakimoto, {\sl On $E_{10}$},
 in: ``Differential geometrical methods  in theoretical physics'',
 eds. K.~Bleuler and M.~Werner, (Kluwer, 1988).


\bibitem{Kleinschmidt:2006dy}
  A.~Kleinschmidt, H.~Nicolai and J.~Palmkvist,
  {\sl $K(E_9)$ from $K(E_{10})$},
  JHEP {\bf 0706} (2007) 051
  [arXiv:hep-th/0611314].

\bibitem{Damour:2007bd}
  T.~Damour and H.~Nicolai,
  {\sl Symmetries, Singularities and the De-Emergence of Space},
  Int.\, J.\ Mod.\ Phys.\ {\bf D17} (2008) 525
  [arXiv:0705.2643 [hep-th]].

\bibitem{Romans:1985tz}
  L.~J.~Romans,
  {\sl Massive N=2a Supergravity In Ten-Dimensions},
  Phys.\ Lett.\  B {\bf 169} (1986) 374.

\bibitem{Brown:2004jb}
  J.~Brown, O.~J.~Ganor and C.~Helfgott,
  {\sl M-theory and $E_{10}$: Billiards, branes, and imaginary roots},
  JHEP {\bf 0408} (2004) 063
  [arXiv:hep-th/0401053].


\end{thebibliography}
\end{document}